%% file: bare_conf_sp.tex
\documentclass[conference,compsoc]{IEEEtran}
\ifCLASSOPTIONcompsoc
  \usepackage[nocompress]{cite}
\else
  \usepackage{cite}
\fi
\ifCLASSINFOpdf
\else
\fi

\usepackage{amssymb}
\usepackage{pifont}
\usepackage{booktabs}
\usepackage{enumitem}
\usepackage{makecell}
\usepackage{url}
\usepackage{hyperref}
\RequirePackage{xurl}
\usepackage{multirow}

\newcommand{\paragraphb}[1]{\noindent{\bf #1} }

\usepackage{array}
\usepackage[most]{tcolorbox}

\usepackage{subcaption}
\usepackage{alltt}
\usepackage{xcolor}

\usepackage{tabularx}
\usepackage{booktabs}

\tcbuselibrary{skins,listings,breakable}

\newcommand{\parhead}[1]{\medskip\Parhead{#1}}
\newcommand{\Parhead}[1]{\noindent\textbf{#1}\hskip 0.5em\relax}

\newcounter{finding}
\renewcommand{\thefinding}{\arabic{finding}}
\newtcolorbox{findingbox}{
  colback=gray!10!white,
  colframe=gray!50!black,
  boxrule=0.5pt,
  before upper=\refstepcounter{finding}\textbf{Finding~\thefinding. }\ignorespaces
}

\begin{document}
%
\title{Whose Agent Are You? \\Multi-Layer Fingerprinting and Attribution of Autonomous Web Agents}


\author{\IEEEauthorblockN{Dayeon Kang}
	\IEEEauthorblockA{UMass Amherst\\
		dayeonkang@umass.edu}
	\and
	\IEEEauthorblockN{Hyejun Jeong}
	\IEEEauthorblockA{UMass Amherst \\
		hjeong@umass.edu}
	\and
    \IEEEauthorblockN{Jade Sheffey}
	\IEEEauthorblockA{UMass Amherst\\
		jsheffey@cs.umass.edu}
	\and
	\IEEEauthorblockN{Pubali Datta}
	\IEEEauthorblockA{UMass Amherst\\
		pdatta@umass.edu}
	\and
	\IEEEauthorblockN{Amir Houmansadr}
	\IEEEauthorblockA{UMass Amherst\\
		amir@cs.umass.edu}
    }


%


\maketitle

\begin{abstract}

As AI web agents proliferate,  combining large language models with autonomous, browser-level control, indiscriminate content scraping by web agents has emerged as a privacy and security challenge. Existing defenses, such as \texttt{robots.txt} and active bot-blocking, are insufficient, as they are widely violated and easily circumvented. 
In this work, we demonstrate that \emph{AI web agents can be effectively distinguished from humans and traditional crawlers using a multi-layer fingerprint} based on both network layer characteristics (e.g., TLS, HTTP) and browser interaction behavior.
We implement this mechanism as a programmatic logging framework that can be deployed on a live, instrumented domain.

By analyzing six prominent agent frameworks (AutoGen, Browser Use, Claude, Gemini, ChatGPT, and Skyvern), we uncover latent structural differences in how these systems assemble HTTP requests, establish TLS/HTTP connections, and execute autonomous browser actions. Feeding these multi-layer features into a decision tree classifier, our framework achieves high-fidelity identification (100\% accuracy), successfully isolating distinct agent architectures and differentiating agent traffic from both human browsing baselines and legacy crawlers. Our findings demonstrate that cross-layer agent tracking provides a robust approach to content protection and web security policy enforcement.

\end{abstract}


%
\IEEEpeerreviewmaketitle

\input{sections/intro}
\input{sections/background}

\input{sections/methodology}
\input{sections/analysis}
\input{sections/discussion}

\section{Conclusion}

In this work, we present multi-layer fingerprinting for identifying AI web agents to protect against increasing indiscriminate content scraping that existing defenses, such as \texttt{robots.txt} and active bot-blocking, fail to prevent. We empirically measure and analyze the agents' fingerprints across request timing, TLS, HTTP, and the behavioral layer from six agent frameworks accessing our deployed live testbed. Our analysis reveals that AI web agents leave structurally distinctive fingerprints at each layer, and no single feature is sufficient for agent identification. We feed the extracted features to the Extra Trees classifier, and it achieves 100\% accuracy in classifying agents, crawlers, and humans for the combination of all features. The results demonstrate that cross-layer fingerprinting provides a robust approach to distinguish agent traffic, with each feature layer complementing the others.

\section*{Ethics Considerations}
In our study, we collected human-generated traces, including network traffic and web component interaction event logs, to compare with AI web agent-generated traces as a baseline. Our university’s Institutional Review Board (IRB) determined our human subjects experiment protocol as a review exemption. Participation was voluntary, and participants were informed about the study's purpose and the data being collected before participating. To protect the participants' privacy, we used the authors' personal laptop over a Wi-Fi connection to prevent participants' IP and MAC addresses from being logged by our deployed server. We asked participants to use either fake or real information and did not collect, store, or process any participant PII (Personally Identifiable Information). 

\section*{Artifact Availability}
Our multi-layer fingerprinting framework for AI web agents, including the designed UX testbed and trace logging system, is available at: \url{https://github.com/SPIN-UMass/AI-agent-fingerprint}



%

\bibliographystyle{IEEEtran}
\bibliography{main}



\input{sections/appendix}

\end{document}

%% file: sections/intro.tex
\section{Introduction}

Beyond the passive text generation capabilities of earlier language models, Large Language Models (LLMs) have evolved into active agents capable of independently navigating the Internet, interacting with webpage elements through clicks and form submissions, and completing complex tasks on behalf of users. As LLM-powered web agents are increasingly deployed in production systems~\cite{masterofcode150Agent}, an emerging concern is their scraping of protected content without authorization. This shift from passive to active AI has fueled rapid AI adoption from both industries and individuals~\cite{hbsWhosAdopting}. 

Website operators have long relied on the \texttt{robots.txt} file (RFC 9309~\cite{rfc9309}) to signal to crawlers which pages should and should not be accessed. 
However, recent work~\cite{10.1145/3719027.3765063, 10.1145/3730567.3764471, 10.1145/3730567.3732913} has documented widespread \texttt{robots.txt} violations by LLM bots, and Cloudflare has observed Perplexity bypassing these directives altogether by impersonating legitimate users through user-agent header spoofing~\cite{cloudflarePerplexityUsing}. Websites have responded with active countermeasures, such as serving CAPTCHAs or misleading content to detected bots~\cite{cloudflareTrappingMisbehaving}, and routing traffic through reverse proxies like Cloudflare. Yet these approaches carry significant limitations: active blocking is inherently binary, cannot replace \texttt{robots.txt} across all crawler types, and is adopted by only 2\% of the top 10K websites~\cite{10.1145/3730567.3732913}. Neither mechanism offers a complete solution. What is needed is a more fundamental capability: the ability to reliably identify and attribute AI agents through their behavior.

Various fingerprinting research studies have focused on identifying targets from the traces users leave when accessing or browsing websites. Website fingerprinting~\cite{sun2002statistical, panchenko2016website, hayes2016k} analyzes ingress and egress traffic timing and burst patterns to infer which websites a user visits, while browser fingerprinting~\cite{vastel2018fp, eckersley2010unique, pugliese2020long} detects bots by collecting device and browser attributes, such as browser type and version, operating system, screen resolution, and installed fonts. Behavioral fingerprinting~\cite{banse2012tracking, gu2017novel, herrmann2013behavior, yang2010web, obendorf2007web, deusser2020browsing} extends this further by capturing biometric signals --- keystroke dynamics, mouse movement, and navigation habits --- to identify individual users.
AI web agents introduce a distinct attribution setting that does not fit neatly into these existing categories. Unlike traditional bots, which often follow predefined automation routines, AI agents are autonomous bots that automatically perceive web content, make decisions, and interact with web components to complete tasks. 
The combination of bot-generated network traces and agent-specific interaction patterns that are affected by human prompts makes identifying and distinguishing heterogeneous agent systems a different challenge from conventional bot detection or individual-user fingerprinting. 
Building on recent attention to detecting AI agents from network traffic~\cite{webfingerprint2026}, we extend fingerprinting beyond traditional single-layer analysis to multi-class attribution among heterogeneous AI agents.


In this work, we measure and characterize a diverse set of AI web agents by fingerprinting their behavior across temporal traffic patterns, multiple network layers (TLS, HTTP), and through browser-level interaction patterns.
Inspired by some web agents \cite{skyvern2026} that parse web pages based on screenshots through visual cues, and others \cite{browser_use2024, wu2023autogen} that do so by rendering an HTML file, we design five scenarios to capture agents' UX behavior: (1) submitting input and clicking a fake or legitimate button, (2) scrolling through the page about the argument, clicking the button, and seeing a counterargument appear, (3) hovering over a button embedded in a panel, (4) answering a short question via a button that uses the wrong answer's ID instead of displayed text, and (5) interacting with a button that produces delayed feedback. These scenarios are deployed on a live website, 
 which we have operated and instrumented fully under our control since October 1.

Our logging infrastructure, hosted on AWS EC2, captures every incoming packet's IP address, timestamp, TLS parameters, and HTTP headers, while the front-end records user interactions (e.g., clicks, inputs, scrolls, and mouse movements) at 1 ms resolution. To generate agent traces, we issue prompts that direct each agent to explore the site and make autonomous decisions (see Figure~\ref{fig:task_prompt} in Appendix~\ref{appen:measurement}). We evaluate agents across a range of architectures: GPT-5.4-mini as a backbone for three open-source frameworks (AutoGen~\cite{wu2023autogen}, Skyvern~\cite{skyvern2026}, and BrowserUse~\cite{browser_use2024}) 
and computer-use agents from Claude~\cite{anthropic2026computeruse}, Gemini~\cite{google2026computeruse}, and ChatGPT~\cite{chatgpt-work}. As a baseline comparison, we collect human- and traditional crawler-generated (Nutch~\cite{apacheApacheNutch}, Heritrix~\cite{githubGitHubInternetarchiveheritrix3}, and Scrapy~\cite{scrapyScrapyOpen}) traces under the same experimental instructions as the agents. 

With this dataset in hand, we isolate traces by prompt issuance time, trial, and agent type, and then analyze each agent's unique features across multiple layers from the network traffic temporal patterns, TLS records, and HTTP requests to behavioral patterns when interacting with web components. By analyzing agents' traces across four analytical dimensions, we discovered that temporal patterns of requests enable evident identification among different agents, and AI-assembled HTTP headers contain bot-indicative signals, such as \texttt{Sec-Fetch} rule violations or 100\% headless Chrome usage. AutoGen and Browser Use have distinct network-level features; they contain bot-indicative signals in different HTTP header fields\textemdash User Agent inconsistency appears in AutoGen, and \texttt{Sec-Fetch} logic violation appears in Browser Use. 
Gemini and Claude show almost identical packet configuration strategies, but they are identifiable via unique web browsing behavior (mouse movement and keydown) caused by their decision-making and interaction strategies. 
Although behavioral-layer signals are influenced by webpage content and experimental conditions, the six agents exhibit distinct interaction patterns that provide useful discriminative information for agent attribution.
For robust agent identification, four different cross-layer features work complementarily. 

The distinctive characteristics we identify across agents inform a feature extraction pipeline spanning all observed layers; these features are then aggregated as inputs to a decision tree classifier designed to uncover latent behavioral signatures that distinguish one agent from another. With the simple classifier, the combination of four features identifies agent types with 100\% accuracy. For real-time agent identification as the requests are incoming, network-level features outperform in the early stage before behavioral signals are fully captured, and behavioral features supplement the discriminability afterwards. After three requests are received, the classifier can detect which agent, crawler, or human is browsing the website with more than 60\% F1 score. 

Our contributions are summarized as follows:

\begin{itemize} 
    \item We propose a multi-layer attribution framework that characterizes AI web agents through complementary features spanning request timing, TLS fingerprints, HTTP headers, and browser interaction behaviors. 
    \item We build and deploy an instrumented website featuring five controlled UX scenarios and collect interaction traces from six AI web agents, human users, and traditional crawlers, establishing a dataset for systematic analysis under controlled yet diverse scenarios. 
    \item We empirically demonstrate that the collected features enable 100\% identification accuracy across six AI web agent types, human users, and traditional crawlers, exhibiting the discriminative power of cross-layer behavioral features under our experimental setup. 
\end{itemize}

%% file: sections/background.tex
\section{Background \& Related Work}~\label{sec:background}

\subsection{AI Agents}
AI agents extend LLMs beyond passive text generation into systems that can pursue goals through iterative interaction with external environments.
At the center of an agent is a backbone LLM, or a set of LLMs, that serves as the agent's decision-making core \cite{bazinska2026breaking}.
Specifically, the LLM is embedded in a perception-action loop: given a user instruction, the agent observes the current state, reasons about the next step, selects an action, executes that action through tools or interfaces, and incorporates the feedback into subsequent decisions \cite{yao2023react, wu2023autogen, wang2023voyager}.

Agents are often categorized by the environments they operate in and the tasks they support.
Web agents, including AutoGen \cite{wu2023autogen}, Browser Use \cite{browser_use2024}, Skyvern \cite{skyvern2026}, and OpenAI Operator \cite{openai2025operator}, browse websites and perform online tasks through screenshots or actions such as searching, clicking, scrolling, and form filling.
Computer-Use Agents (CUAs), such as OmniTool \cite{microsoft2025omnitool}, Claude's CUA \cite{anthropic2026computeruse}, and Gemini's CUA \cite{google2026computeruse}, interact with the computer GUI via mouse and keyboard to automate desktop tasks.
Coding agents, such as SWE-Agent \cite{yang2024sweagent}, Cursor \cite{cursor2026}, OpenAI Codex \cite{openai2026codex}, Claude Code \cite{anthropic2026claudecode}, and Devin \cite{cognition2024devin}, write, debug, and refactor code by editing files, running commands, and incorporating execution feedback.
Research agents, such as GPT Researcher \cite{elovic2023gptresearcher}, OpenAI's Deep Research \cite{openai2025introducingdeepresearch}, and Gemini's Deep Research \cite{google2025geminideepresearch}, coordinate retrieval, browsing, and synthesis across multiple information sources.
Although these categories are not mutually exclusive, they illustrate the diversity of agentic systems and the range of execution traces they produce.
We focus on web agents in this work.

\subsection{Different Types of Web Agents}
We distinguish web agents by how LLMs are used for browser automation: how the agent reads the webpage, parses page content into a model-consumable representation, identifies actionable elements, and selects the next browser action \cite{stockl2025ai}.
In this context, ``reading'' and ``understanding'' a webpage refer to the agent's perception pipeline, where rendered pixels, HTML/DOM structure, accessibility metadata, or parsed UI elements are converted into an observation that is fed to the LLM \cite{skyvern2025readsweb}.
This distinction matters because different designs make page information available to the model at different times.
For example, a vision-based agent may need to scroll and re-observe before it can select a button below the current viewport, whereas a DOM- or accessibility-tree-based agent may consider the same button earlier if the serialized page representation includes off-screen elements.
For hybrid agents, we classify them by their primary parsing logic.

\paragraphb{Vision-Based Web Agents.}
Vision-based web agents read webpages primarily through rendered screenshots and select actions through visual grounding.
They act through mouse and keyboard operations such as clicking, typing, and scrolling.
For example, OpenAI Operator is powered by the CUA, which combines GPT-4o's visual capabilities with reasoning and interacts with webpages through browser actions \cite{openai2025operator,openai2025cua}.
Similarly, Claude's and Gemini's CUAs expose screenshots to the model and execute generated mouse or keyboard actions in the browser or desktop environment \cite{anthropic2026computeruse,google2026computeruse}.
Skyvern is also vision-first: its public description emphasizes using screenshots, computer vision, and LLM reasoning to identify webpage elements such as buttons, forms, links, and tables, rather than relying on brittle DOM selectors \cite{skyvern2025readsweb}.
These agents approximate human-visible browsing because their observations are grounded in rendered screen content.

\paragraphb{DOM-Based Web Agents.}
DOM-based web agents read webpages through structured page representations such as HTML, DOM trees, or browser state.
Instead of relying primarily on visual screenshots, these agents parse the underlying page structure and choose actions that are executed through browser automation.
For example, Browser Use navigates webpages through Chromium via CDP, parses HTML, and repeatedly queries an LLM to decide the next action \cite{browser_use2024}.
This representation can make textual and structural page content easier for the model to process, but it can also change the agent's information boundary by exposing content that may be hidden, off-screen, or visually de-emphasized to human users.

\paragraphb{Accessibility-Tree-Based Web Agents.}
Accessibility-tree-based agents read webpages through semantic roles, labels, and element identifiers rather than raw pixels or the full DOM.
The accessibility tree provides a compact, interaction-oriented representation of the page that preserves elements such as buttons, text boxes, links, headings, and form fields.
AutoGen's WebSurfer is a representative example: in Magentic-One, WebSurfer operates a Chromium browser and relies on the browser accessibility tree together with a set of marks that prompt it to perform webpage actions \cite{wu2023autogen}.
This interface gives the model a structured semantic view of the page and can ground actions to interactable elements, while abstracting away some visual layout and styling information.

\subsection{Client Fingerprinting}

\Parhead{Network-Level Fingerprinting.}
Network-level fingerprinting identifies a client's \emph{networking stack} by characterizing metadata from packets sent by the client at various layers.
Hus\'ak et al.\ first showed that clients can be passively identified from the cipher suite lists offered in their SSL/TLS handshakes, building a dictionary that maps handshake parameters to client types~\cite{husak2016https}.
This approach was later popularized by JA3, which hashes the \texttt{ClientHello} version, cipher suites, extensions, and elliptic-curve parameters into a compact, shareable fingerprint~\cite{althouse2019ja3}; it was also refined by JA4+, which sorts the extension list so that fingerprints remain stable under Chrome's \texttt{ClientHello} extension randomization~\cite{althouse2023ja4}.
Frolov et al.~\cite{frolov2019tls} measured TLS fingerprintability across billions of real-world connections and released the \texttt{utls} library for parsing and mimicking arbitrary \texttt{ClientHello} messages, while Anderson et al.\ showed that the mapping from TLS fingerprint to application is many-to-one~\cite{anderson2019tls}, motivating the fusion of additional signal layers.
Above TLS, Akamai researchers demonstrated that HTTP/2 clients can be passively fingerprinted from their connection preludes, i.e., \texttt{SETTINGS} frame values, the initial \texttt{WINDOW\_UPDATE} increment, \texttt{PRIORITY} frames, and pseudo-header order~\cite{segal2017http2}; we adopt this fingerprint format in our measurement framework (\S\ref{sec:methodology}).
These signals underpin server-side bot detection: Li et al.\ exposed large-scale bot activity claiming browser User-Agents while being built on simple HTTP libraries, flagged through inconsistencies between the advertised User-Agent and the observed TLS handshake~\cite{li2021goodbot}.
Recent preprints have begun to apply network-level fingerprinting to AI-driven clients: Jarad and Bicakci detect malicious bots from JA4 handshake features, but only as a binary bot-versus-human decision~\cite{jarad2026handshakes}, and AgentPrint identifies LLM-based agents from encrypted traffic timing and volume patterns observed on the user--agent link~\cite{zhang2025agentprint}.
In contrast, we perform multi-class attribution of \emph{specific} web agents from the server side, fusing TLS, HTTP/2, header, and timing fingerprints with browser-level behavioral signals.

\paragraphb{Behavioral Fingerprinting.}
Behavioral fingerprinting identifies users by capturing habitual patterns of how users interact with web content. Previous works~\cite{banse2012tracking, gu2017novel, herrmann2013behavior, yang2010web, obendorf2007web, deusser2020browsing} have focused on low-level interaction signals, such as keystroke dynamics and mouse movement, as individual identification features. Extended works investigate broader navigation habits, including page visit sequences, dwell times, and scrolling behavior, showing that browsing style alone can be an identifier. Behavioral fingerprints have traditionally been regarded as requiring long-term observation, typically over 24 hours, to accumulate sufficient signal for reliable identification~\cite{banse2012tracking, gu2017novel, herrmann2013behavior, kirchler2016tracked, vassio2017users}. 

Crichton et al.~\cite{crichton2025rethinking} provide a large-scale analysis of behavioral fingerprinting over 150,000 users across two years and find that it is highly unique and temporally stable, evolving slowly over months. They demonstrate that a user loses 78--85\% of anonymity within the first 60 seconds of a new browsing session. Furthermore, they show that combining behavioral and browser fingerprinting outperforms either method individually, achieving an F1 score of 0.869 across 100,000 users. 

As an attempt to use the behavioral fingerprinting method in detecting AI web agents, Wang et al.~\cite{wang2026fpagentfingerprintingaibrowsing} analyze the typing, scrolling, and mouse behavior of agents. They also show that combining behavioral and browser fingerprinting outperforms either approach alone, which is consistent with the findings of Crichton et al. However, this work adopts existing browser fingerprinting as a complement to behavioral signals, without accounting for the fundamentally different nature of AI agents as LLM-controlled autonomous systems. Our work proposes an agent-specific multi-layer fingerprinting framework that targets the unique structural properties of agent systems.


\begin{figure*}
    \centering
    \includegraphics[width=\linewidth]{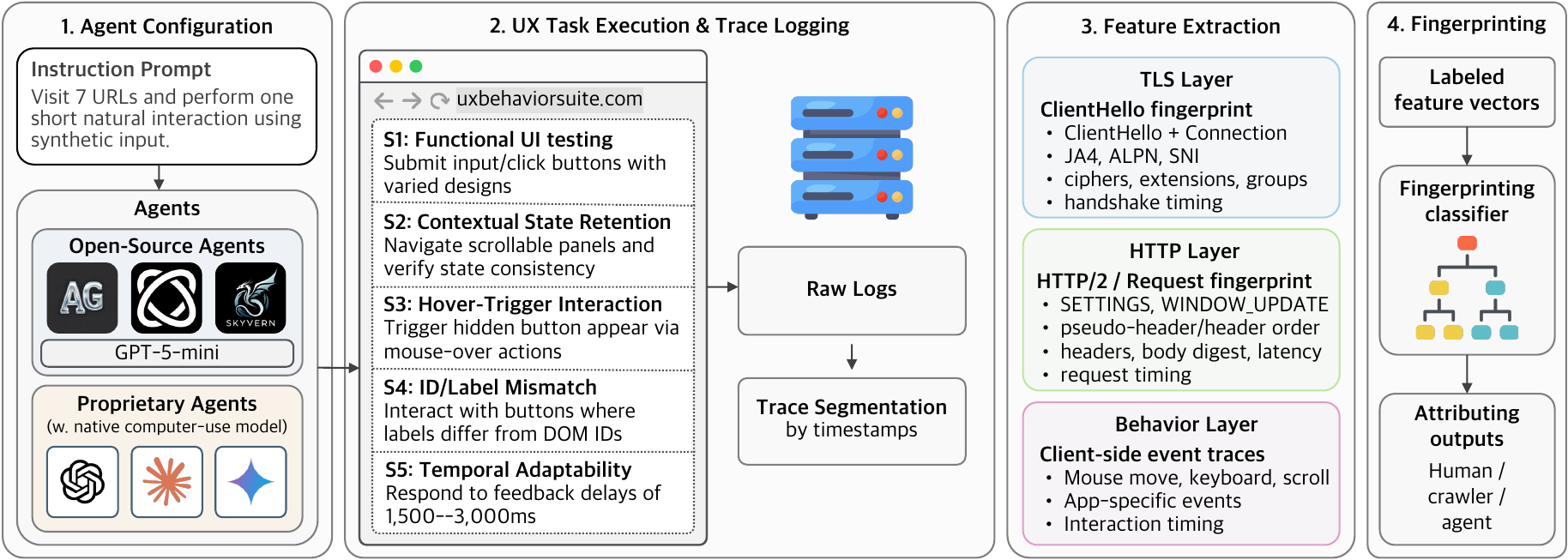}
    \caption{\textbf{Overview of MARK.} Our \textbf{M}ulti-layer \textbf{A}gent fingerprinting framewo\textbf{RK} (MARK) consists of four stages: 
    (1) configuring agents with a common web-interaction instruction, 
    (2) collecting raw network and client-side traces as each agent visits URLs and performs controlled UX tasks, 
    (3) extracting TLS, HTTP, and behavioral features from the segmented traces, and 
    (4) using the resulting feature vectors to attribute traces to humans, crawlers, or web agents.}
    \label{fig:overview}
\end{figure*}

%% file: sections/methodology.tex



\section{MARK: Multi-layer Agent fingerprinting framewoRK}\label{sec:methodology} 


We present MARK, a multi-layer agent fingerprinting framework that passively records and analyzes traces left by web agents at four observation layers: request timing, TLS, HTTP, and in-browser behavior. As Figure~\ref{fig:overview} illustrates, MARK operates at the website end without requiring any client-side instrumentation, records traces, and extracts feature vectors from four levels that feed into an agent identification classifier. Identifying agents with a classifier is addressed in Section~\ref{sec:eval}.

\subsection{UX Task and Webpage Design}\label{sec:ux-task}
Different web agents interact with webpages differently.
Motivated by this observation, we design five user-web interaction scenarios that can surface discrepancies in how agents perceive and interact with webpages.
We focus on how web agents access implicit or hidden information and interact with webpages spatially (e.g., mouse movements, clicks, scrolls) or temporally (e.g., timing delays, responsiveness).

\textbf{Scenario 1} targets agents' ability to distinguish functional from non-functional UI elements. 
We create three pages, each with an input field and a button next to it. 
Two pages (i.e., v1 and v3) have the exact same layout and the same text button label (i.e., submit), but only one successfully submits the input.
The other page (v2) uses a button label with an icon instead of text with a correct submission target.
\textbf{Scenario 2} probes whether agents retain contextual state across interactions.
A scrollable panel appears at the center of the page, with a clickable button at the bottom of the panel. Upon clicking the button, another passage appears. 
\textbf{Scenario 3} examines hover-dependent interaction. 
A clickable button appears only when the user's mouse cursor hovers over the containing panel. 
This tests whether agents discover the button through mouse movement or access it directly from the underlying HTML without hovering.
\textbf{Scenario 4} introduces a semantic mismatch.
A common-sense quiz presents answer buttons whose HTML label IDs do not match their visible text.
It tests whether agents select answers based on the visible button text or the underlying HTML labels.
\textbf{Scenario 5} measures temporal adaptability by introducing a randomized delay of 1,500--3,000~ms between a button click and the appearance of a confirmation popup. This scenario distinguishes agents that wait for feedback from those that proceed without it.

\subsection{Trace Logging System}

\begin{table*}[t]
\centering
\scriptsize
\caption{\textbf{Logged fields for fingerprint configuration across three layers.}}
\label{tab:logged-fields}
\setlength{\tabcolsep}{3.5pt}
\begin{tabularx}{\linewidth}{@{}p{1.75cm}p{2.9cm}>{\raggedright\arraybackslash}X@{}}
\toprule
\textbf{Layer} & \textbf{Category} & \textbf{Fields} \\
\midrule
\textbf{TLS}
  & Connection
  & Negotiated version, cipher suite, SNI \\

  & Fingerprint hashes
  & JA3 \cite{althouse2019ja3}, JA4 \cite{althouse2023ja4} \\

  & ClientHello fields
  & Offered cipher suites, ALPN offer list, supported groups, EC point formats, signature schemes, supported TLS versions, extension count (from JA4) \\
  
\midrule
\textbf{HTTP}
  & HTTP/2 prelude
  & \texttt{SETTINGS} frame values (\texttt{HEADER\_TABLE\_SIZE}, \texttt{ENABLE\_PUSH}, \texttt{MAX\_CONCURRENT\_STREAMS}, \texttt{INITIAL\_WINDOW\_SIZE}, \texttt{MAX\_FRAME\_SIZE}, \texttt{MAX\_HEADER\_LIST\_SIZE}), 
  \texttt{WINDOW\_UPDATE} increment, 
  pseudo-header order (\texttt{:method}, \texttt{:scheme}, \texttt{:authority}, \texttt{:path})
  \\

  & Request metadata
  & Method, path, protocol, header order, body size, SHA-256 body digest, server-side latency \\

  & Automation headers
  & \texttt{User-Agent}, \texttt{Sec-Ch-Ua*}, \texttt{Sec-Fetch-*}, \texttt{Accept-*}, \texttt{Referer}, \texttt{Cookie} \\

\midrule
\textbf{Behavior}
  & Keyboard \& Mouse events 
  & \texttt{keydown}, \texttt{keyup}, \texttt{mousemove}, \texttt{mousedown}, \texttt{mouseup}, \texttt{click}, \texttt{dblclick}, \texttt{contextmenu} \\

  & Input \& focus
  & \texttt{input}, \texttt{touchstart}, \texttt{touchend}, \texttt{change}, \texttt{focus}, \texttt{blur}, \texttt{selectionchange} \\

  & Page events
  & \texttt{scroll}, \texttt{resize}, \texttt{load}, \texttt{beforeunload}, \texttt{pagehide}, \texttt{visibilitychange} \\

  & Application events
  & Correct/incorrect submission, hover-reveal trigger, delayed-popup arrival \\

  & Event metadata
  & Wall-clock timestamp, per-tab session ID, CSS-selector target descriptor \\

\bottomrule
\end{tabularx}
\end{table*}

The testbed is hosted at the registered domain on an AWS EC2 instance and runs as a single Go service that terminates TLS on port 443 (with a Let's Encrypt certificate~\cite{letsencryptAboutLets}). The service simultaneously captures three layers of fingerprint information for every visit: TLS, HTTP, and in-browser behavior.

We build on the \texttt{fingerproxy} library so that the raw \texttt{ClientHello} record and the negotiated connection state are exposed to our handlers. We record TLS connection states, compute fingerprint hashes, and re-parse the \texttt{ClientHello} through \texttt{utls}. \texttt{ClientHello} fields preserve sufficient detail to reconstruct the hashes and to detect client-library substitutions that would collide under a single hash.
To log the HTTP/2 connections, we extract the Akamai-style fingerprint \cite{segal2017http2} from the prelude and additionally log the request metadata for HTTP/1.1 and HTTP/2. We record per-request server-side latency from the request and response timestamps. 
All entries are appended as JSONL into a daily-rotated \texttt{requests-YYYY-MM-DD.jsonl} file.
Each scenario page loads a lightweight script, \texttt{logger.js}, that subscribes to event profiles. 
For \texttt{mousemove}, it is throttled at $16~\text{ms}$ to cap payload size at the $60~\text{Hz}$ refresh rate while still resolving sub-second cursor trajectories. Each event is tagged with a wall-clock timestamp computed from \texttt{performance.now()} for sub-millisecond precision. 
Scenario pages also emit named application events through a small \texttt{window.logEvent} API, which lets us align in-browser actions with the scenario-specific ground truth. The detailed recorded entries of three layers are summarized in Table~\ref{tab:logged-fields}.

A key design choice is that the logger issues \emph{no mid-session HTTP requests}. Events are buffered in memory and flushed only at end-of-page (\texttt{beforeunload}/\texttt{pagehide}), when the tab becomes hidden, or when the buffer exceeds 5{,}000 events, in each case via \texttt{navigator.sendBeacon}. This avoids contaminating the TLS and HTTP/2 traces with telemetry-induced requests that would otherwise be indistinguishable from agent-driven traffic and would skew per-page request counts. Each flush carries an incrementing sequence number and a shared session identifier, so the server reassembles multi-flush sessions on the receiving end. The server's \texttt{/collect} endpoint persists each batch verbatim into a daily \texttt{interactions-YYYY-MM-DD.jsonl} file, keeping the behavior log structurally parallel to the request log and joinable by source IP, session identifier, and timestamp during analysis.

\subsection{Feature Extraction}\label{sec:feature-extraction}


After the request and event traces are collected, we segregate them based on the timestamps of agent browsing. 
Before feature extraction, sendBeacon traffic emitted by the browser-side logger is removed from request records to preserve only accessing agent-generated traffic. From each trace, we extract four feature vectors from different extractors independently.

Temporal features are derived from the timestamp sequence of HTTP requests. The extractor calculates the Inter-Request Interval (IRI) distribution (mean, standard deviation, and median) and request rate in Hz. Since the variance of the request rate differs by agent, the Coefficient of Variation (CV) of the request rate is computed after all traces of the same agent are aggregated. Additionally, the extractor captures page dwell times and subresource burst patterns. In total, 9 temporal feature vectors reflect the agent's request pacing strategy.

The TLS extractor parses the TLS and HTTP/2 fields of each request and aggregates them per trial into thirteen feature groups (36 features in total). Post-Quantum key exchange is captured as a fraction of connections offering X25519MLKEM768, together with the fraction in which it is offered but not preferred first. From JA4, we use the extension count from the JA4\_a segment and the fraction of connections deviating from dominant browser's JA4; JA3 and JA4 are further summarized as per-trial diversity. The TLS~1.3 cipher offer order is encoded as an ordinal mode, and the counts of offered cipher suites, supported groups, and signature schemes are averaged per trial. 
HTTP/2 PRIORITY frame data is parsed via the Akamai H2 fingerprint format, extracting the \texttt{WINDOW\_UPDATE} size, pseudo-header order, initial window size, presence of \texttt{MAX\_CONCURRENT\_STREAMS}, the presence of stream-3 and stream-5 PRIORITY frames, and the stream-5 priority weight. GREASE usage across cipher suites, curve groups, and supported versions is recorded, along with a set of agent-exclusive ClientHello/HTTP/2 values (e.g., 12 signature schemes, more than 20 cipher suites).

The HTTP header extractor operates at request level, enumerating seven feature groups (total 35 features): a hash of the navigation header ordering sequence, the \texttt{Sec-Fetch-*} rule violation, client coherence between \texttt{Sec-Ch-Ua} and the \texttt{User-Agent} header field, per-header presence flags for standard browser headers, \texttt{User-Agent} family, OS, and version parsed from the UA string, \texttt{Priority} header urgency value and its correctness according to the request role (navigate, subresource, or async fetch), and the non-standard custom header names. At the trace level, these per-request signals are aggregated as rates into scalar and string values. 

From the separately stored user event log, the behavioral extractor captures browser events triggered by user actions (i.e., clicks, keydowns, mouse movements, scrolls, inputs, and navigation). Features are consist of ten groups, total 108: session-level counts and event rates, inter-event interval distribution statistics, event-type ratios (e.g., mousemove-to-keydown ratio), timing and reaction features including input latency and key hold duration, mouse-move behavior (i.e., total trajectory length, click coordination, and click targets), keyboard input behavior (i.e., Enter key ratio, special-key usage (e.g., Control, Alt, Command, Function, Shift) usage, and key diversity), scroll depth and count, page-level structure including dwell times and a marking of which scenario pages were visited, task completion signals from application-level events, and navigation style measured by tab-click transitions between pages. Additionally, the extractor computes per-page and per-event-type proportions. 

\section{Measurement Setup}


With our framework, MARK, we measure AI web agent traces systematically across multiple layers to perform a comprehensive analysis of collected fingerprints. 

\subsection{Task Execution Prompt for Agents}\label{sec:measurement-prompt} 

We run the same task execution prompt for each agent.
The prompt asks the agent to visit seven target webpages in a fixed order, perform one short and natural interaction on each page, and then immediately proceed to the next page.
We design the prompt to avoid common execution failures, such as stalling on a single page, asking the user for clarification, or merely describing the page without interacting with it.
To keep traces comparable across agents, the prompt also instructs agents to use synthetic inputs, avoid real personal information, and complete each page after one meaningful interaction sequence.
Figure~\ref{fig:task_prompt} in Appendix~\ref{appen:measurement} shows the full task instruction.

\subsection{Agents and Backbone Models}
We measure web-agent behavior by running each agent on the same controlled browser-interaction task with a shared instruction prompt. 
For all open-source agents, we use \texttt{GPT-5.4-mini} as the backbone model for AutoGen, Browser-use, and Skyvern. 
For proprietary agents, we use each system's native computer-use model: \texttt{GPT-6.1 Sol} with medium thinking mode for the ChatGPT agent, \texttt{claude-sonnet-5} and \texttt{computer-use-2025-11-24} for the Claude agent, and \texttt{gemini-3.8-flash} for the Gemini agent. 
We keep the backbone model fixed to fingerprint agents, not models, while treating proprietary systems as bundled agent-model implementations.

Since web-agent trajectories are inherently noisy and non-deterministic \cite{akkil2026emergence}, we run each agent 30 times to capture consistent behavioral patterns in page inspection, action selection, timing, and interaction order.
We use these repeated traces to support more generalizable comparisons across agents, even for consistent and systemic browsing actions of web agents in the downstream fingerprinting analysis.

When agents access the test-bed website through the registered domain name and conduct the tasks, our fingerprinting framework, MARK, records fingerprints generated from three different layers: TLS, HTTP, and browser interaction level. Two files are stored on the deployed server: \texttt{requests-YYYY-MM-DD.jsonl}, which contains incoming TLS and HTTP traffic, and \texttt{interactions-YYYY-MM-DD.jsonl}, which records user-generated interaction events at the frontend.


%% file: sections/analysis.tex
\section{Attributions of Web Agent Fingerprinting}\label{sec:observation}







We analyze multi-layer fingerprints collected from six AI web agents to identify which features are discriminative for the agent. Across three levels\textemdash request-level temporal distributions, network-layer protocol signals, and browser interaction behavior\textemdash we characterize what makes each agent distinct and assess which feature space is sufficient to isolate agents.

\subsection{Requests-Level Fingerprints}

\begin{figure}[t]
    \centering
    \begin{subfigure}[b]{0.49\columnwidth}
        \centering
        \includegraphics[width=\textwidth]{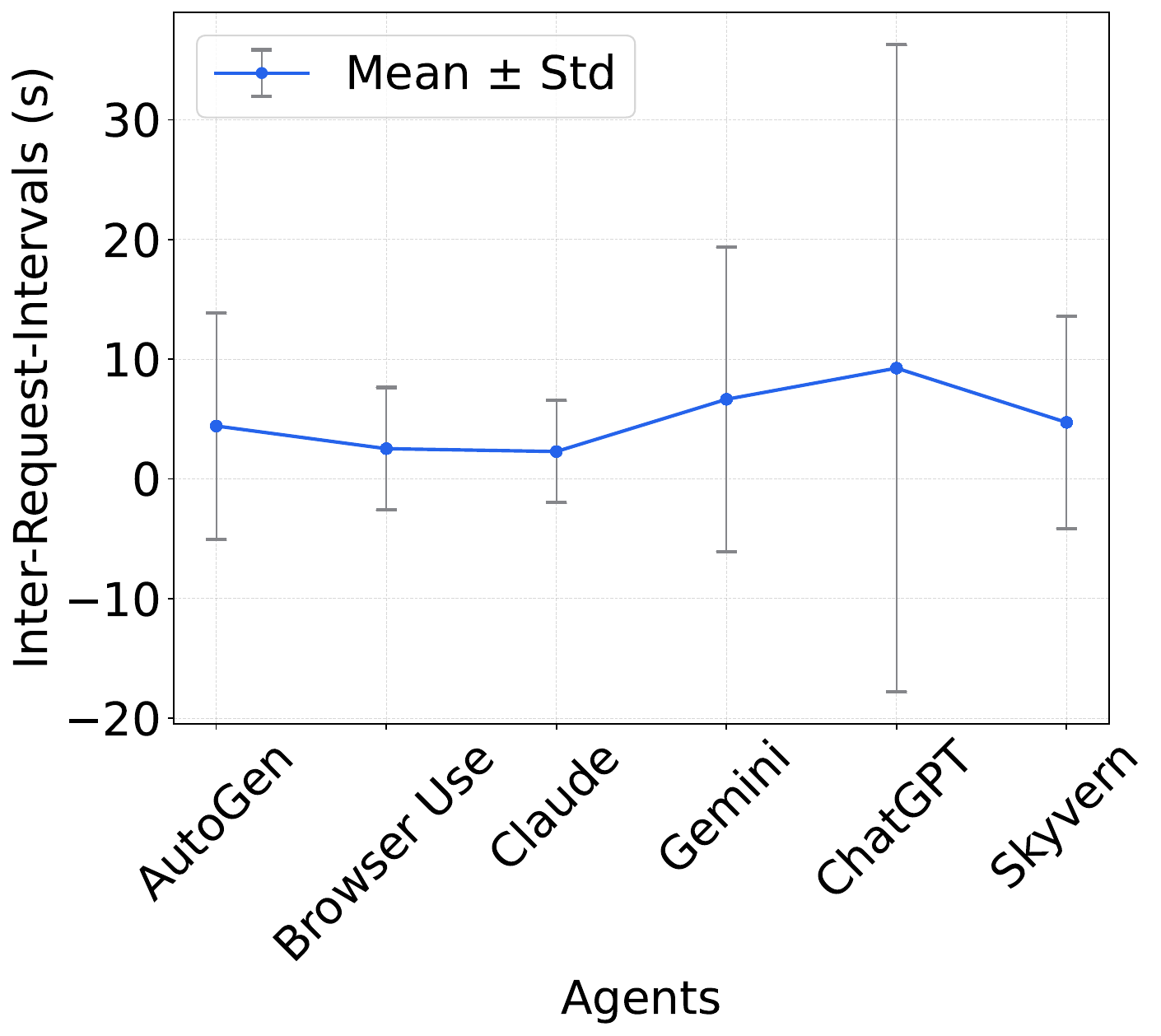}
        \caption{Inter-Request-Intervals}
        \label{fig:iri}
    \end{subfigure}
    \begin{subfigure}[b]{0.49\columnwidth}
        \centering
        \includegraphics[width=\textwidth]{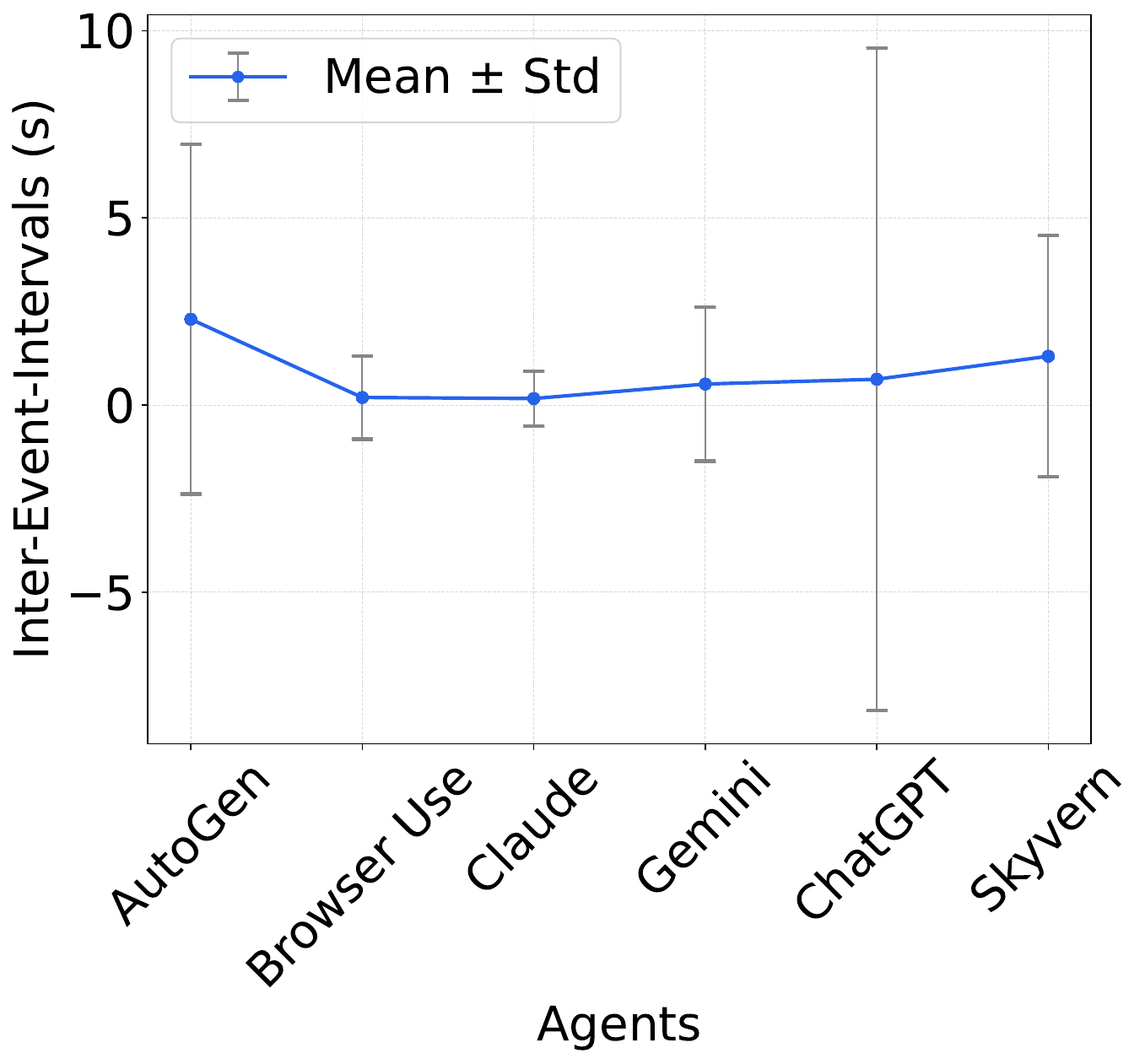}
        \caption{Inter-Event-Intervals}
        \label{fig:iei}
    \end{subfigure}
    
    \caption{\textbf{Comparison of request timing (Inter-Request-Intervals and Inter-Event-Intervals).} A similar tendency between IRI and IEI indicates the pacing strategies of agents.}
    \label{fig:iri-iei}
\end{figure}

\begin{figure}[h]
  \centering
  \includegraphics[width=\linewidth]{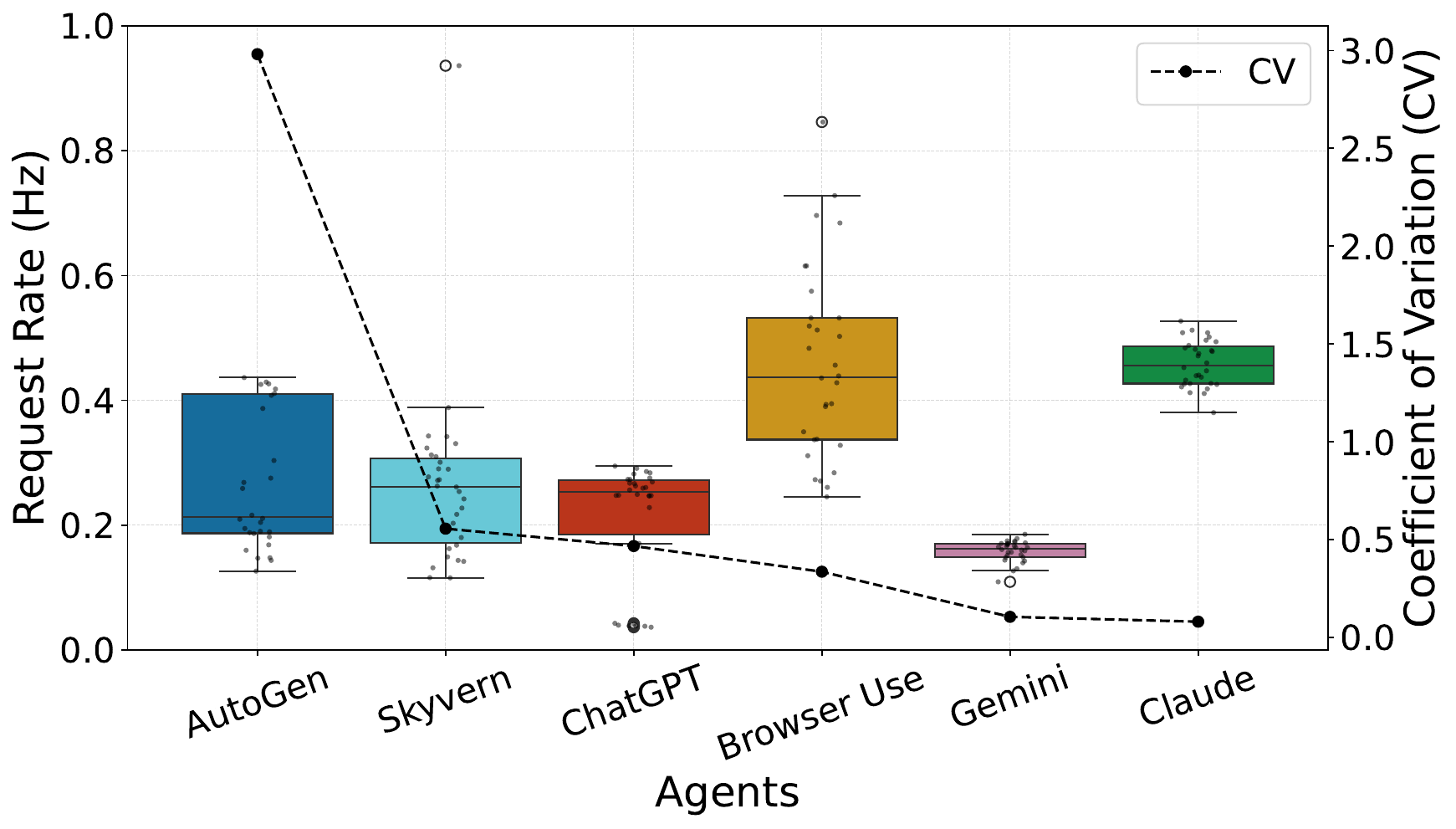}
  \caption{\textbf{Comparison results of the request rate sent by the agent and its Coefficient of Variation.} This feature shows the stability and pacing of each agent.}
  \label{fig:req-rate}
\end{figure}

Borrowing from traditional website fingerprinting methods that analyze the temporal distribution of network traffic packets, we examine temporal distributions at the HTTPS request level, including Inter-Request Intervals (IRI) and Inter-Event Intervals (IEI) within one session and request rate across trials. As shown in Figure~\ref{fig:iri}, ChatGPT exhibits the largest variation in inter-request intervals. This stems from two sources: within a trial, requests are emitted in short bursts separated by long pauses, and across trials, 7 of 30 trials took roughly 700~seconds to complete compared to the rest. These prolonged idle periods likely reflect waiting on the agent side (e.g., for confirmation or extended reasoning). Meanwhile, its short mean inter-event interval (Figure~\ref{fig:iei}) indicates that once ChatGPT acts, it executes interaction events in rapid succession. Combining the request rate and Coefficient of Variation (CV) that provides stability of request patterns in Figure~\ref{fig:req-rate}, Claude shows fast and systemically consistent traffic patterns even in repeated trials, while Gemini is slow but consistently performs tasks with even intervals rather than long idle time. Browser Use is fast, similar to Claude, in one session, but the speed differs by trial. Interestingly, AutoGen shows the highest inter-event intervals, and this can be assumed to be due to AutoGen's structure of calling the LLM for each action.

These observations suggest that request-level temporal distributions are influenced more by the design of the agent system than by the underlying agent type itself. In particular, the temporal features, reflecting the agent’s pacing strategy and action-taking process, may drift depending on the task or website. Motivated by this, we next investigate more stable features that are less sensitive to task variation, including TLS and HTTP header characteristics.

\begin{findingbox}
    Claude behaves most consistently but sends requests quickly, while Gemini conducts tasks slowly but consistently. ChatGPT and AutoGen exhibit high variability with a large variation in IRI and IEI, and CV, respectively. 
\end{findingbox}

\subsection{Protocol-Level Fingerprints}

\begin{table*}[ht]
\scriptsize
\centering
\setlength{\tabcolsep}{5pt}
\renewcommand{\arraystretch}{1.3}
\caption{%
  \textbf{Agent-discriminative TLS and HTTP/2 features.}
  CC20 denotes \texttt{CHACHA20\_POLY1305\_SHA256}, and AES128/256 denote \texttt{AES\_128/256\_GCM}. \textbf{Bold} marks discriminative values.} 
\label{tab:tls_fingerprints}

\begin{tabular}{@{}l cc c r l@{}}
\toprule
& \multicolumn{2}{c}{Cipher preference} & \multicolumn{1}{c}{HTTP/2 \textsc{priority}} & \multicolumn{2}{c}{Deviation from dominant ClientHello} \\
\cmidrule(lr){2-3}\cmidrule(lr){4-4}\cmidrule(lr){5-6}
Agent & \makecell[c]{Cipher offer order} & \makecell[c]{Negotiated cipher} & \makecell[c]{Stream-5 weight\\(3rd request)} & \makecell[r]{Off-dominant\\JA4 (\%)} & \makecell[l]{Agent-exclusive value\\(coverage)} \\
\midrule
\textbf{AutoGen} & CC20$\to$AES128$\to$AES256 (93\%) & CC20 (93\%) & 256 & \textbf{7.2} & cipher list\,=\,AES256$\to$CC20$\to$AES128, 31 ciphers (3.4\%) \\
\textbf{Browser Use} & \textbf{AES128$\to$AES256$\to$CC20} & \textbf{AES128} & \textbf{220} & 0.6 & JA4\,=\,\texttt{t13d131100\_f57a46bbacb6\_e57\ldots} (0.2\%) \\
\textbf{Claude} & CC20$\to$AES128$\to$AES256 & CC20 & 256  & 0 & \textit{none} \\
\textbf{Gemini} & CC20$\to$AES128$\to$AES256 & CC20 & 256  & 0.8 & \textit{none} \\
\textbf{ChatGPT} & CC20$\to$AES128$\to$AES256 (98\%) & CC20 (98\%) & 256  & 2.1 & curves\,=\,\texttt{29,23,24,4588} (1.3\%) \\
\textbf{Skyvern} & \textbf{AES128$\to$AES256$\to$CC20 (99\%)} & \textbf{AES128} & 256 & \textbf{14} & \textbf{pseudo-header\,=\,\texttt{a,m,p,s} (13\%)} \\
\bottomrule
\end{tabular}
\end{table*}

\Parhead{TLS/H2 fingerprint.}
First, we analyze the TLS/H2 fingerprint of incoming traffic from the website end. As we experimented with the default browser, Chromium-based, the key configurations of TLS ClientHello are almost identical. Therefore, we focus on three distinctive configurations: cipher offer order, HTTP/2 PRIORITY weight, and off-dominant JA4 (Table~\ref{tab:tls_fingerprints}). The most powerful signal to distinguish agents in the TLS layer is TLS 1.3 cipher preference. AutoGen, Claude, Gemini, and ChatGPT offer ciphers in CC20$\rightarrow$AES128$\rightarrow$AES256 order, but Browser Use and Skyvern offer AES128$\rightarrow$AES256$\rightarrow$CC20. Even with the same Chromium browser, Chromium is known to switch the priority between ChaCha20 and AES-GCM depending on whether hardware AES acceleration is enabled in the runtime environment. This difference is likely to reflect the host environment in which the agent executes (cloud VM, container, CPU type) rather than the agent software itself. This cipher order and negotiated cipher appears consistently in 93–100\% of agents, which indicates this feature stably distinguish agents into two groups.



Most of the connections' JA4 hashes start with \texttt{t13d1516h2}, so we focused on the off-domain ClientHello, which is the subclient trace, to observe whether the agent uses the other HTTP client as well. 13\% of Skyvern connections use pseudo-header \texttt{a,m,p,s}, and this value appears only in Skyvern, suggesting a separate HTTP client other than Chrome sends some requests directly. AutoGent also exhibits 7.2\% of off-dominant JA4 with 3.4\% of legacy 31 ciphers, indicating that it is possibly a non-browser client (e.g., OpenSSL-based library) that sends requests. Browser Use, Gemini, and Claude present almost pure Chrome TLS fingerprints. 

\begin{findingbox}
    Cipher offer order distinguishes agents into two groups (i.e., negotiated cipher as CC20 and AES 128) consistently, which is likely to be affected by the agent's hosting environment. 
\end{findingbox}

The dominant values of window update for flow control and pseudo-header order in the Akamai H2 fingerprint format are consistent between agents. However, the weight of stream 5 and its percentage in traffic are unique. 
HTTP/2 multiplexes multiple requests over a single TCP connection as streams. Each stream gets an odd ID assigned sequentially by the client: the first request is stream 1, the second is stream 3, the third is stream 5, and so on.
The original HTTP/2 spec (RFC 7540~\cite{rfc7540}) included a priority tree mechanism where each stream could declare a dependency on another stream and a relative weight (1–256). With the weight, the browser informs the server which HTTP requests are more important and allocates more connection resources. Chrome implements this as a specific priority tree, and it sends at connection start via \texttt{PRIORITY} frames.
Weight 256, the maximum, means Chrome is telling the server to treat that stream as the highest priority relative to other streams.
Weight 220, which appears in Browser Use, suggests a slightly different Chrome build or OS-level scheduler configuration.
In recent Chrome versions, most \texttt{PRIORITY} frame transmission has been discontinued as RFC 9218~\cite{rfc9218} (HTTP/2 Extensible Priority) discontinues the previous priority tree due to the preference for a simpler header-based method. However, the agent still contains priority frames, and certain weights consistently applied for each agent make them distinguishable. 

\begin{findingbox}
    Browser Use has a uniquely identifiable TLS/H2 fingerprint.
    It is the only agent with an AES128-negotiated cipher and a distinct stream-5 priority weight of 220.
\end{findingbox}

\begin{table*}[htbp]
  \centering
  \caption{\textbf{HTTP-header fingerprint across six AI web agents.} HTTP-header discriminates agents via non-human browsing signals. 
  Red indicates anomalous or bot-indicative behavior, and * indicates inconsistent value across traces.}
  \label{tab:http-header-fingerprint}

  \resizebox{\linewidth}{!}{%
  \begin{tabular}{lcccccc}
    \toprule
    \textbf{Feature}
      & {\textbf{AutoGen}}
      & {\textbf{Browser Use}}
      & {\textbf{Claude}}
      & {\textbf{Gemini}}
      & {\textbf{ChatGPT}}
      & {\textbf{Skyvern}} \\
    \midrule

  UA family (mode) & edge\textsuperscript{*} & headless-chrome\textsuperscript{*} & headless-chrome & headless-chrome & headless-chrome & chrome \\

    HeadlessChrome in \texttt{Sec-Ch-Ua} (\%)
      & \textcolor{red}{93} & 0 & \textcolor{red}{100} & \textcolor{red}{100} & \textcolor{red}{100} & 0 \\

    \texttt{Accept-Language} present (\%)
      & \textcolor{red}{3} & 100 & \textcolor{red}{0} & \textcolor{red}{0} & \textcolor{red}{0} & 100 \\


    \texttt{Sec-Fetch} logic violation (\%)
      & 0 & \textcolor{red}{89} & 0 & 0 & 0 & 0 \\

    UA vs.\ Sec-CH-UA incoherence (\%)
      & \textcolor{red}{93} & 0 & 0 & 0 & 0 & 0 \\

    Unique nav. header orderings
      & 199/200 & 209/210 & 207/210 & 207/210 & 209/210 & 206/210 \\

    Non-standard headers
      & \texttt{pragma} & \textit{none} & \textit{none} & \textit{none} & \textit{none} & \textit{none}\\

    \bottomrule
  \end{tabular}%
  }
\end{table*}

\Parhead{HTTP header fingerprint.}
Next, we analyze the HTTP request headers that web agents sent (Table~\ref{tab:http-header-fingerprint}). In the HTTP \texttt{User-Agent} header, the browser family is inconsistent in agents, but Browser Use, Claude, Gemini, and ChatGPT consistently have headless Chrome in the \texttt{User-Agent} header. We additionally investigate the \texttt{Sec-Ch-Ua} header. Interestingly, Browser Use has \texttt{HeadlessChrome} value as a UA family with \texttt{Sec-Ch-Ua: "Not;A=Brand";v="24", "Chromium";v="128""}, which is coherent as Chrome, but not consistent. Also, 93\% of AutoGen requests contain HeadlessChrome, but the dominant value of the UA family is Edge. The significant incoherence of \texttt{User-Agent} and \texttt{Sec-Ch-Ua} is a unique characteristic of web agents.
AutoGen, Claude, Gemini, and ChatGPT HTTP headers show strong headless and automation bot-indicative behavior because nearly all headers of HTTP requests omit \texttt{Accept-Language}. 

Notably, Browser Use is the most uniquely identifiable agent due to the impossible \texttt{Sec-Fetch} semantics. 
Browser Use superficially mimics human browsing, omitting HeadlessChrome in \texttt{Sec-Ch-Ua} and including 100\% of \texttt{Accept-Language}, while violating fetch metadata semantics. 
We discover that the request's \texttt{Sec-Fetch-*} headers (Mode, Dest, Site, User) in Browser Use have mismatched values that a real-world user accessing through the browser would not have. \texttt{Sec-Fetch-Mode} represents how the request was initiated (navigate, no-cors, cors, same-origin), \texttt{Sec-Fetch-Dest} represents what the request is for (document, script, image, empty, etc.), \texttt{Sec-Fetch-Site} represents the relationship to origin (same-origin, cross-site, none), and \texttt{Sec-Fetch-User} represents whether it was a user gesture (\texttt{?1} only for navigation clicks). 
We flag a request as a logic violation when its \texttt{Sec-Fetch-*} headers contain semantically impossible combinations that no real browser would produce. Specifically, a violation occurs when the \texttt{Sec-Fetch-Mode} and \texttt{Sec-Fetch-Dest} pair is unknown or invalid, when the \texttt{Sec-Fetch-Site} value does not match the request context\textemdash for example, a \texttt{(no-cors, image)} sub-resource request carrying \texttt{Sec-Fetch-Site: none} is impossible because embedded image loads always originate from within a page. Additional violations occur when a \texttt{Sec-Fetch-Mode: navigate} request omits \texttt{Sec-Fetch-User: ?1} despite navigations always being user-initiated, and when a CORS fetch request carries a non-empty \texttt{Sec-Fetch-Dest} value rather than \texttt{empty}, which is what real browsers normally send for XHR/fetch requests.

\begin{findingbox}
    Browser Use does not follow the \texttt{Sec-Fetch} rule in 89\% of HTTP requests, which is highly suspicious behavior, as neither bot nor human browsing behavior shows this. 
\end{findingbox}

In real browsers with human access, the header order is deterministic. However, some automatic frameworks break header order stability because headers are assembled dynamically or middleware injects fields asynchronously. Remarkably, we observe that every AI web agent has high variability in the navigation header order, which implies that the agent dynamically reorders headers based on inferred user intent and adapts available actions to the current task context as part of the reasoning system.

AutoGen introduces non-standard headers \texttt{pragma}. These custom headers are a strong fingerprint feature because they are easy to detect at the server side and directly expose the agent identity. The HTTP request header of Skyvern is the closest to human browsing except for the high variance in header ordering. 

AI web agents are increasingly accurately mimicking human browsing HTTP semantics to evade bot detection, but still leak automatic AI bot-specific fingerprints through cross-header inconsistencies: \texttt{Sec-Fetch} rule violations and high dynamic header ordering behavior.

\begin{findingbox}
    AutoGen, Claude, Gemini, and ChatGPT are strongly headless and exhibit bot-indicative behavior, while Skyvern shows near-real-browser behavior with realistic HTTP header formatting.
\end{findingbox}

\begin{findingbox}
    Claude, Gemini, and ChatGPT exhibit highly similar network-layer fingerprints, suggesting that they may share common infrastructure or underlying system components.
\end{findingbox}


\subsection{Behavior-Level Fingerprints}

\begin{figure}[h]
  \centering
  \includegraphics[width=\linewidth]{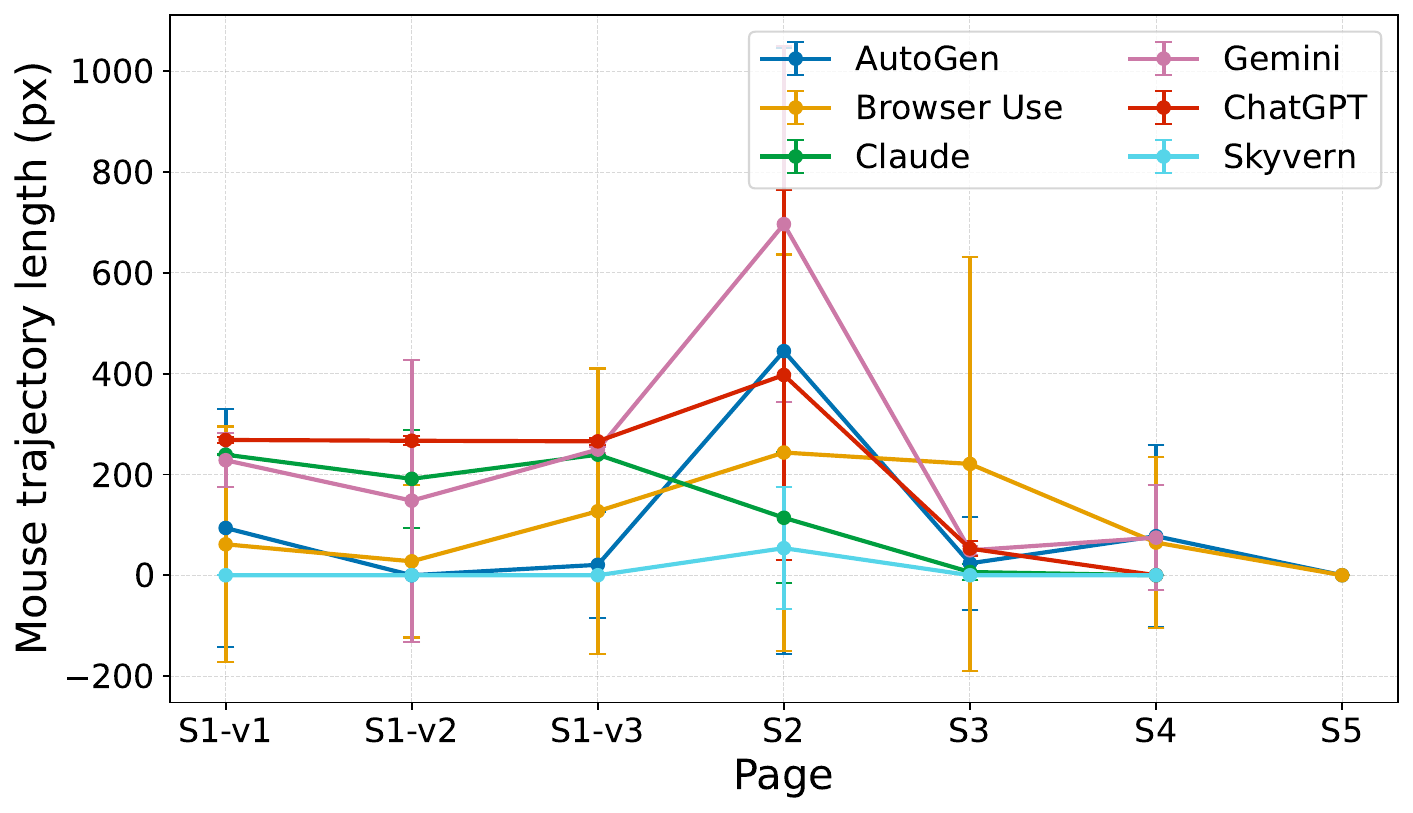}
  \caption{\textbf{Mean and standard deviation values of mouse trajectory length.} It implies agent type and strategy for human-mimicking behavior.}
  \label{fig:mouse-traj}
\end{figure}

\begin{figure}[h]
  \centering
  \includegraphics[width=\linewidth]{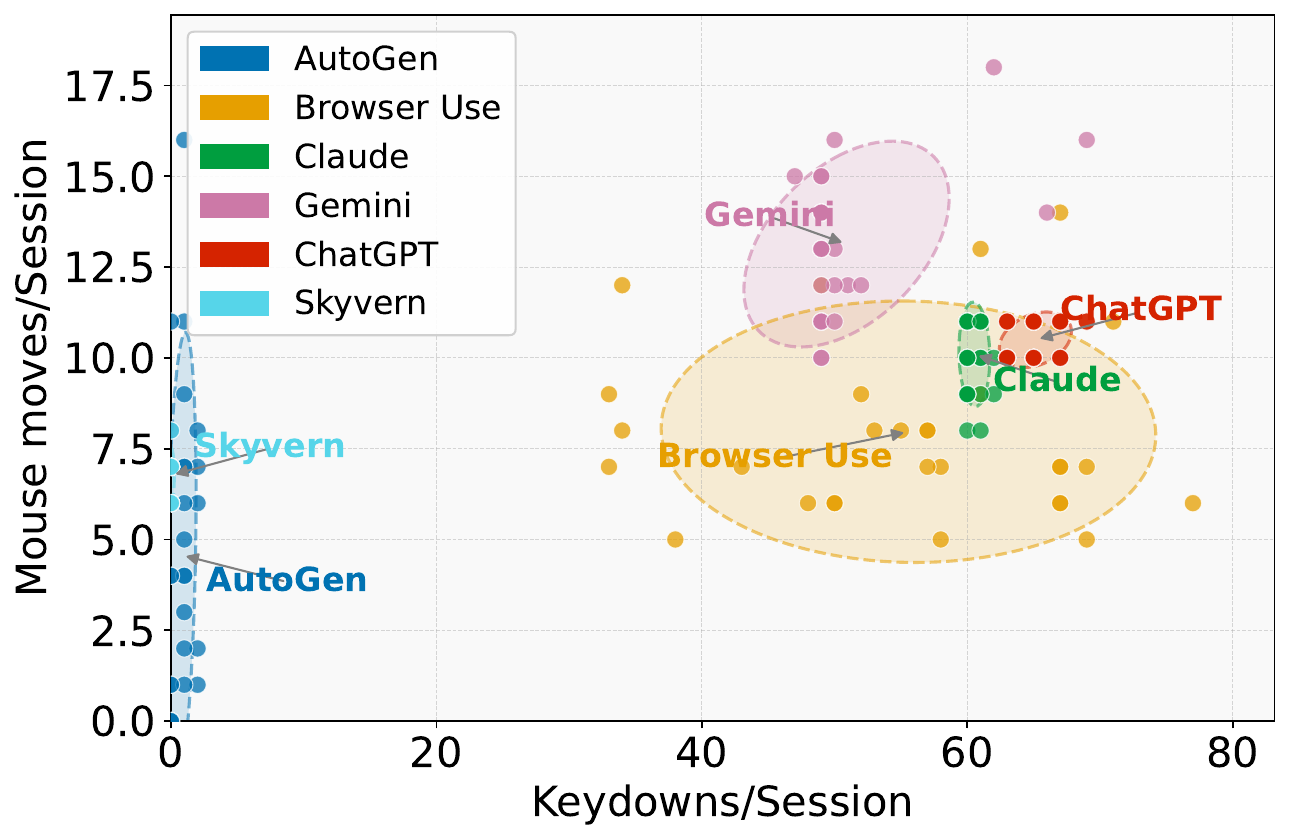}
  \caption{\textbf{The key downs and mouse movements per session clustered regions across agents.} Agents generally use keyboard controls rather than mouse movements, except for Skyvern, which presents human-like behavior.}
  \label{fig:keydown-mousemove}
\end{figure}



Assuming that different types of agents affect decision-making and web-browsing strategies, and that these differences would be reflected in web component interaction behavior, we analyze behavioral features from short web component interaction logs at the website end. 
Interestingly, in all 30 trials, Claude, Gemini, ChatGPT, and Skyvern skipped Scenario 5, a delayed feedback of button click. Claude, Gemini, ChatGPT, and Skyvern are all plan-first (Plan-and-Execute) agents that enumerate the URL list before beginning execution. Scenario 5, last in the navigation, was thus the most susceptible to being dropped, while AutoGen and Browser Use are closer to reactive step-by-step executors.

We track the agent's mouse coordination with a 1~ms interval and calculate the total mouse trajectory length for each web component interaction scenario (Figure~\ref{fig:mouse-traj}). Notably, Skyvern (vision- and DOM-based agent~\cite{how-skyvern-works}) shows the shortest mouse movement because it takes a screenshot of the page, extracts an abridged tree of interactable DOM elements (e.g., input fields, buttons, and links), and lets the LLM decide the next action. AutoGen also moves the mouse less, except for scenario 2, which has a scrolling and click-and-popup panel, similar to Skyvern because AutoGen uses a vision- and DOM-based method that takes screenshots, extracts the interactive elements, and sends a set-of-mark (SoM) screenshot, which draws bounding boxes, to the LLM model~\cite{how-autogen-websurfer-works}. Since Skyvern and AutoGen adopt element-based Playwright, their keydown is nearly zero, as shown in Figure~\ref{fig:keydown-mousemove}, even though they submitted the input in scenario 1. 

Claude, Gemini, and ChatGPT exhibit similar and more than 150px mouse movement in scenario 1 and around 40-70 keydowns. This is because they are vision-based~\cite{notteDoesBrowser, tinyfishWhatBrowser}, showing the actual mouse movement from one click point to the next and entering each character when submitting the input. In Figure~\ref{fig:keydown-mousemove}, Browser Use has the fluctuation in the number of inputs and the biggest ellipse size among the six agents, suggesting its behavior varies a lot across sessions. Possible reasons for such variation are: when the input field is filled incorrectly and re-filled because the element's index becomes stale, which is a common pattern of a DOM index-based agent~\cite{githubLogicalArchitecture, techstackupsBrowserClaude}, and a change in the task completion process. Combined with the consistent and fast request rate of Browser Use, we estimate that it has a stable agent loop timing, but the LLM may make a different judgment every time. In particular, the small ellipse size of Skyvern, Claude, and ChatGPT indicates how their component interaction behavior is deterministic.

\begin{findingbox}
    Behavioral telemetry can distinguish agents when network fingerprints are similar.
    Claude, Gemini, and ChatGPT exhibit almost identical network fingerprints, but they are separable through deterministic and subtly different mouse moves and keydown patterns. 
\end{findingbox}


Observing the other event profile of agents, Skyvern did not issue keydown events while input is submitted, which indicates it directly inputs values without keystrokes. In scenario 3---the button appears when the mouse hovers over the panel---only AutoGen and Browser Use clicked the button.
Integrating all behavioral features observed from the webpage side, we complementarily analyze each agent from a different perspective. 
With a coarse-grained, session-level keydown and mouse-move interaction, Gemini, ChatGPT, and Claude are isolated, but Browser Use's scatter area overlaps with them. The fine-grained page-level event profile contributes to the fact that some agents with unique events discriminate the scatter-area-overlapped agents---for example, Skyvern shows contrast in input and keydown events, and Browser Use clicks the button in scenario 3. In our controlled UX scenarios, six targeted agents can be fully identified.


\begin{findingbox}
    Agents' interaction types (e.g., primarily vision-based, DOM-based) lead to distinct mouse movement and keyboard input patterns and differences in behavioral determinism. Also, some agents exhibit a unique event profile due to their underlying structure. 
\end{findingbox}

\section{Fingerprint-Based Agent Identification in Practice}\label{sec:eval}


As we showed in Section~\ref{sec:observation}, our cross-layer fingerprints can distinguish different AI web agents. Using these observed fingerprint characterizations, we design a decision tree classifier to identify the accessing user, including agents, humans, and traditional crawlers, with collected traces in practice.

\subsection{Baseline Data Collection}

\Parhead{Human-generated Traces.}
To obtain a human baseline, we recruit 30 human participants and log their interactions with the same web interaction suite used for agents. 
After participants consent, we give them three instructions before they begin the task. 
First, some pages (i.e., the three different versions of our scenario 1) include an email input field, designed to make the subscription button scenario more realistic than a standalone button. 
Participants are told that they can enter either a real or a fake email address, and that the input will not be submitted, registered, or used to subscribe them to anything. 
This clarification is necessary because some participants are otherwise reluctant to enter any information into an unfamiliar newsletter form. 

Second, participants are instructed not to revisit pages they have already visited. 
This mirrors the agent protocol, in which each agent is prompted to visit the seven URLs sequentially and interact with each page once. 
We do not ask agents to retry until they succeed because our goal is to collect behavioral traces from a first-pass interaction, rather than to measure whether the agent can complete a fixed page-specific task. 
A retry-until-success protocol would require defining success criteria for each page in advance, which would turn the experiment into a task-completion benchmark and make the traces less reflective of natural interaction behavior. Applying the same one-pass constraint to human participants, therefore, makes the human and agent traces procedurally comparable. 

Third, after completing all seven interactions, participants are asked to close the browser tab, giving each session a clear endpoint and preventing post-task behavior from being included in the trace.

When participants are ready to begin, we direct them to visit the testbed. 
We record the interaction trace, including the start and end times of each session, and monitor whether participants adhere to the interaction protocol.

\Parhead{Crawler-generated Traces.}
To contrast AI agents against conventional automated clients, we collect a baseline from three widely used traditional web crawlers, chosen for their distinct network stacks: \texttt{Scrapy}~2.14.1 (Python/Twisted, built on pyOpenSSL/OpenSSL), \texttt{Heritrix}~3.15.0 (the Internet Archive's archival crawler, Java~17/JSSE), and \texttt{Apache Nutch}~1.22 (Java, using its OkHttp-based HTTP protocol plugin over JSSE). These span the dominant TLS and HTTP client libraries used by non-browser automation, so that any signal shared across them is characteristic of legacy crawlers in general rather than of a single implementation.

Each crawler is pointed at the same instrumented testbed, seeded at the site root, and left to traverse the site using its standard link-following behavior while honoring \texttt{robots.txt}, under the tool's default fetch configuration. Mirroring the 30-trial protocol used for the agents, we execute 30 independent crawl sessions per crawler. Because the testbed is a live domain that also receives background scanner traffic and all of our crawl traffic egresses from a single IP, we run the sessions strictly one at a time and sequentially, separated by a fixed idle gap so that no two crawl windows overlap. We attribute the server-side records of each session using its wall-clock UTC window and cross-check that every attributed record originates from our single measurement egress IP; across all 90 sessions, this yields clean, non-overlapping attribution with no foreign-IP contamination.

The resulting traces are request-only. Unlike the AI agents, none of the crawlers executes JavaScript, so \texttt{logger.js} never fires, and they emit no interaction (behavioral) telemetry. All three communicate exclusively over HTTP/1.0 or HTTP/1.1 and never negotiate HTTP/2, leaving the Akamai HTTP/2 fingerprint empty. Each crawler nonetheless presents a stable, distinct TLS fingerprint that is constant across all 30 of its sessions\textemdash a single JA3 hash for Scrapy and Nutch, while Heritrix consistently emits two distinct JSSE \texttt{ClientHello}s\textemdash giving us a deterministic, non-LLM reference point against which the agent and human fingerprints are compared.

\subsection{Agent Identification}

With the collected traces from six AI web agents, humans, and three web crawlers, we extract four feature vectors from four different layers as in Section~\ref{sec:feature-extraction}. To verify that features provide sufficient evidence to identify agents/humans/crawlers, we design a classifier.

\Parhead{Preprocessing.}
Before feeding features into a classifier, we preprocess extracted features into a fully numeric form with trial-level feature matrices for a decision tree-based classifier.
Temporal, TLS, and behavioral features are already numeric, so we remove zero-variance columns whose values are constant across all trials and replace missing values for some trials with the median value. 
We exclude the coefficient of variation (CV) of the request rate because it is computed by aggregating traces from the same agent and could introduce information from the held-out fold if computed before cross-validation, potentially inflating classification performance.
Since the HTTP header features contain the names of non-standard headers, we perform multi-hot encoding of the custom headers before dropping zero-variance values. In addition to the numeric format feature preprocessing strategy, missing values of \texttt{Sec-Fetch-Site}, \texttt{Sec-Fetch-Mode}, User-Agent family, and User-Agent OS are substituted with unknown, and missing values of \texttt{Priority} header become zero. 
Finally, agent labels are integer-encoded for each preprocessed feature, and labels are uniformly shared across feature vectors. 

\Parhead{Decision Tree Classifier.}
We selected an Extra Trees (Extremely Randomized Trees) as a classifier, which is an ensemble of fully randomized decision trees with faster training and performs well when features have varying scales. Considering the feature configuration\textemdash the mix of numeric and encoded categorical features, and the moderate size of the dataset\textemdash Extra Trees is well-suited, and the extracted features have the ability to identify the agent without a sophisticated deep learning model. 

We train an Extra Trees classifier separately on each of the five feature sets: temporal, TLS/H2, HTTP header, behavioral, and four combined features. All models are trained with 200 estimators and a fixed random seed for reproducibility. To obtain unbiased performance estimates, we apply cross-validation. In each of the 30 folds, all feature vector rows are used as the test set once, and the model is trained on the remaining 29 trials of each agent. After cross-validation, a final Extra Trees classifier is trained on all feature vector rows for each feature type. With the trained models, we evaluate and report the classifier's agent identification performance from various angles, as well as the detection of human and legacy crawlers at the same time. 

\subsection{Identification Results}

\begin{table}[ht]
\centering
\caption{\textbf{Cross-validation results of Extra Trees classifier by feature type.} Temporal features identify agents with fully observed traces.}
\label{tab:feature_type_results}
\begin{tabular}{lrrrr}
\toprule
Feature Type & Accuracy & Precision & Recall & F1 \\
\midrule
Temporal & 0.857 & 0.857 & 0.857 & 0.852 \\
TLS & 0.767 & 0.807 & 0.767 & 0.744 \\
HTTP & 0.820 & 0.811 & 0.820 & 0.790 \\
Behavioral & 1.000 & 1.000 & 1.000 & 1.000 \\
Combined & 1.000 & 1.000 & 1.000 & 1.000 \\
\bottomrule
\end{tabular}
\end{table}

\begin{table}[ht]
\centering
\scriptsize
\caption{\textbf{Per-agent F1 score by feature type.} Four features supplement the identification of agents/humans/crawlers. (We omitted behavioral features of crawlers because they vary by JavaScript implementation)}
\label{tab:agent_f1_results}
\begin{tabular}{lrrrrr}
\toprule
Agent & Temporal & TLS & HTTP & Behavioral & Combined \\
\midrule
AutoGen & 0.716 & 0.565 & 0.983 & 1.000 & 1.000 \\
Browser Use & 0.750 & 0.984 & 1.000 & 1.000 & 1.000 \\
ChatGPT & 0.949 & 0.462 & 0.439 & 1.000 & 1.000 \\
Claude & 0.903 & 0.484 & 0.556 & 1.000 & 1.000 \\
Gemini & 0.900 & 0.000 & 0.000 & 1.000 & 1.000 \\
Skyvern & 0.793 & 0.984 & 0.966 & 1.000 & 1.000 \\
Human & 0.542 & 0.966 & 0.952 & 1.000 & 1.000 \\
Heritrix & 1.000 & 1.000 & 1.000 & -- & -- \\
Nutch & 0.984 & 1.000 & 1.000 & -- & -- \\
Scrapy & 0.984 & 1.000 & 1.000 & -- & -- \\
\bottomrule
\end{tabular}
\end{table}


We classify feature vectors of AI web agents, humans, and crawlers using the Extra Trees classifier.
Since the crawlers do not interact with web components using JavaScript and depend on the script designed by developers, behavioral features are omitted. 
We also skipped the combined feature set of crawlers for a fair comparison. 

\subsubsection{Full-Trial Identification Performance}
Table~\ref{tab:feature_type_results} presents cross-validation results for each feature vector set, averaging the evaluation results of 30 folds. Temporal features, widely used in traditional website fingerprinting, achieve an accuracy of 85.7\%, demonstrating that request timing alone provides substantial discriminative information about agent identity. However, this result should be interpreted within the scope of the evaluated scenarios and agents, as temporal patterns may vary with task complexity, network conditions, and agent versions.


TLS features are the weakest fingerprint to identify agents (F1 = 0.744). As in Table~\ref{tab:agent_f1_results}, Skyvern (0.935) and Browser Use (0.984) are relatively well-separated by their distinct negotiated cipher and stream-5 priority weights. 
Aligning with the earlier finding that Claude, Gemini, and ChatGPT share nearly identical network-layer fingerprints in Section~\ref{sec:observation}, Gemini shows notable failures in classification (0.000). However, AutoGen (0.983) and Skyvern (0.966) are almost immediately distinguished even by HTTP headers. As human browsing generates stable HTTP headers, unlike bot-generated headers, human browsing behavior is also easy to identify (0.952).  
Traditional crawlers are the most easily distinguishable targets using only network-layer features due to their fundamental packet configuration difference from browser-based agents. Across request, HTTP, and TLS layers, crawlers are identified with a 100\% F1 score. 

Behavioral features achieve the best classification performance in our controlled scenario, attaining 100\% accuracy, precision, recall, and F1-score, both independently and when combined with other feature types. 
Although Claude, Gemini, and ChatGPT are indistinguishable with the network layer feature, behavioral features can differentiate them. From this finding, we can assume that they adopt different web content browsing and decision-making strategies while they share the same network packet assembling backbone system. 


\begin{figure}[h]
  \centering
  \includegraphics[width=\linewidth]{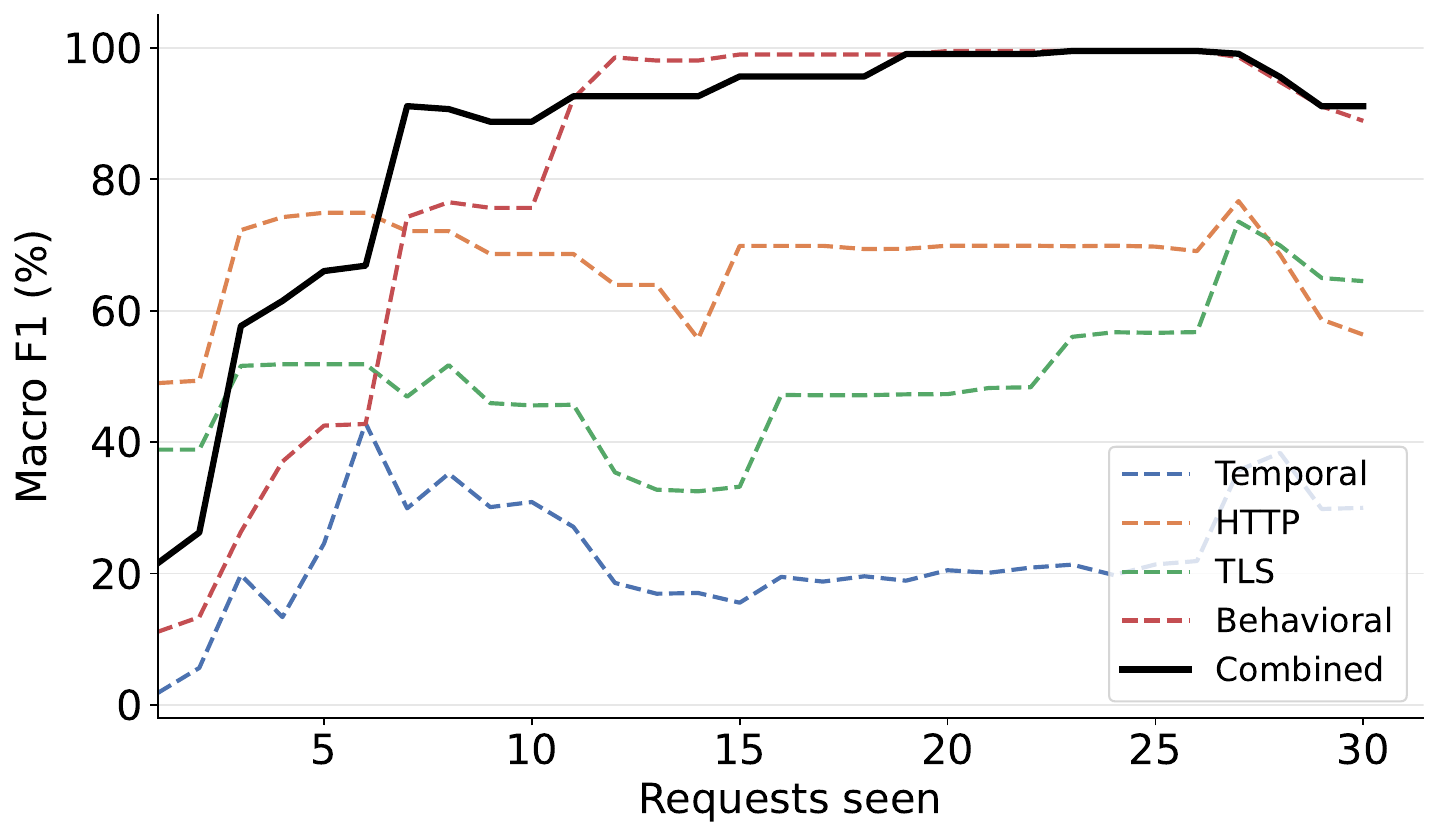}
  \caption{\textbf{Agent identification performance comparison as the number of requests seen at the website end (up to 30 requests).} Protocol-level fingerprints preserve more than 60\% agent discrimination after 3 requests are received, and behavioral features replenish the performance after signals are accumulated.}
  \label{fig:eval_by_request}
\end{figure}


\subsubsection{Early-Request Identification for Real-Time Deployment}
We evaluate the classifier to confirm that it is deployable on the web content hosting server side to protect the content from AI web agents (Figure~\ref{fig:eval_by_request}).
Notably, temporal features perform poorly throughout the entire observation window, showing around 30\% macro F1 score after the first few requests. This is in contrast to the full-trial results reported in Table~\ref{tab:feature_type_results} and reveals a limitation: temporal features such as inter-request/event intervals and request rate require a sufficient volume of requests to obtain a stable distribution estimate. 
TLS and HTTP features both achieve higher performance compared to the behavioral features, which is expected, as only limited event signals are captured in the first few requests. 
The earlier stabilization of HTTP relative to TLS is because HTTP header fingerprints are observable on every request, whereas informative TLS signals such as stream-5 priority weights require multiple concurrent connections for a reliable decision. 
As the agent's browsing behavior is accumulated, the behavioral features increase the accuracy more steeply than the TLS and HTTP features. 

Complementing multiple features with each other throughout the user browsing sessions\textemdash \textit{HTTP and TLS perform well in the early stage, and behavioral features are powerful in the mid-stage discrimination}\textemdash the combined features identify the browsing subject more than 60\% macro F1 score of the time after three requests are received. Figure~\ref{fig:eval_by_request} demonstrates that reliable identification is possible before full requests are observed, leveraging multiple cross-layer features.
To protect web content from agents, the website owner can serve the content selectively after identifying the browsing agent by placing a non-sensitive webpage as a gate. 

\begin{table}[ht]
\centering
\scriptsize
\caption{\textbf{Per-agent F1 score by source-identifying feature ablated feature vector.}}
\label{tab:source-ablating}
\begin{tabular}{lrrrrr}
\toprule
Agent & Temporal & TLS & HTTP & Behavioral & Combined \\
\midrule
AutoGen & 0.716 & 0.512 & 0.652 & 1.000 & 1.000 \\
Browser Use & 0.750 & 0.935 & 0.983 & 1.000 & 1.000 \\
ChatGPT & 0.949 & 0.429 & 0.372 & 1.000 & 1.000 \\
Claude & 0.903 & 0.484 & 0.504 & 1.000 & 1.000 \\
Gemini & 0.900 & 0.000 & 0.000 & 1.000 & 1.000 \\
Skyvern & 0.793 & 0.967 & 0.966 & 1.000 & 1.000 \\
Human & 0.542 & 0.836 & 0.938 & 1.000 & 1.000 \\
Heritrix & 1.000 & 1.000 & 1.000 & -- & -- \\
Nutch & 0.984 & 0.952 & 1.000 & -- & -- \\
Scrapy & 0.984 & 1.000 & 1.000 & -- & -- \\
\midrule
& 0.852 & 0.711 & 0.741 & 1.000 & 1.000 \\
\bottomrule
\end{tabular}
\end{table}

\subsubsection{Separating Agent Attribution from Source Identity}
To assess whether the classifier distinguishes agents based on agent-specific signals rather than source- or environment-specific fingerprints, we evaluate performance after ablating features that may reveal or encode the collection source. In previous identification experiments, we did not use the egress IP directly or the entire ClientHello as a feature vector for classification. However, we used selected ClientHello fields, including cipher suite counts, supported curve counts, signature scheme counts, PQ-related supported groups, and GREASE indicators, as well as source IP diversity.
For source-agnostic identification, we evaluated the classifier after ablating source-identifying features (JA3/JA4 hashes, JA4-derived identity features, TLS sequence counts (cipher, curve, signature), TLS extension count, PQ-related supported groups, source IP diversity, HTTP header-order features, User-Agent-related features, client hints) from both the training and test sets. 
These eliminated features can be influenced by operating system, browser, device, or network path.

As a result, the number of TLS features decreased from 36 to 21, and the number of HTTP features decreased from 37 to 18. This reduction in features led to a slight decrease in F1 scores for TLS and HTTP, from 0.744 to 0.711 and from 0.790 to 0.741, respectively (details in Table~\ref{tab:source-ablating}). The combined F1 score is maintained due to the powerful behavioral features. Despite removing these source-related features, the remaining features retain considerable discriminative power, with the combined F1 score remaining sustained.

%% file: sections/discussion.tex
\section{Performance under Adaptive Adversary}
\parhead{Threat Model.}
Our threat model considers a black-box, simple-adaptive adversary to evade \texttt{robots.txt} files, such as via UA spoofing. The web agents can autonomously interact with the target website, but are not assumed to have knowledge of the defender’s fingerprinting features. 

\parhead{Performance Analysis under Adaptive Adversary.}
We analyze the adversary’s cost of mimicking human behavior, assuming she knows MARK uses four-layer features but cannot observe the exact features or classifier parameters: (1) \textit{behavioral layer} imposes a relatively high cost because there is a limit to reproducing fine-grained human behaviors (e.g., short exploration time, accurate component interaction, human-like page dwell times, etc.)~\cite{guan2026componentbench} (2) \textit{temporal and HTTP-header layers} impose moderate costs, requiring artificial delays and modifications to the automation framework’s request-construction logic, respectively. (3) TLS/H2 features are the most costly, as it requires customizing the browser’s underlying TLS stack. Importantly, spoofing only a subset of layers is unlikely to eliminate MARK’s cross-layer signals. Moreover, prior work~\cite{houmansadr2013parrot} demonstrated that a single imitated session can be detected without correlating multiple flows or performing sophisticated traffic analysis. Thus, unless the adversary collects and closely mimics real human browsing traffic, immature traffic and inconsistencies across layers are likely to remain detectable.

As evaluating MARK under each adversary scenario requires implementation working in real-world observation layer manipulation, we leave evaluation as future work.

\section{Limitations}
We demonstrate that AI web agents leave distinctive traces across request timing, TLS, HTTP, and behavioral layers, enabling reliable classification with features combined from multiple layers. However, our framework has a limitation due to its feature-based identification. As we observed, Claude, Gemini, and ChatGPT agents are indistinguishable because they present almost identical TLS and HTTP fingerprints. If two agents use a fully identical back-end system and decision-making process, both of which shape behavioral signals, our framework will classify them as the same system. More broadly, our framework cannot identify an agent that perfectly mimics human browsing behavior and network traces. 

Our testbed is a controlled environment with ambiguous tasks, such as identical email address submission pages and a mismatch between button ID and displayed text, to induce hesitance and page dwell time. Based on the amount of collected web component interaction events and website design, behavioral fingerprint accuracy may differ. In addition, the temporal features may shift under different task types, website structures, or prompt formulations. 

Since AI web agents are evolving rapidly for efficient and fast task completion, along with underlying LLM models, our classifier's performance may degrade over time as agent behavior changes. Our evaluation does not include a temporal holdout or leave-one-version/configuration-out assessment, which would more rigorously measure robustness against version drift and newly released agent systems. Ongoing feature and classifier updates will be necessary as new models are released. 

Additionally, our dataset was collected under a default browser engines, which are Chrome or Chrome-based. Different browser configurations can introduce variation in network traffic through differences in resource loading strategies, connection management, caching behavior, and request metadata. Therefore, evaluating only one browser environment may limit the diversity of the observed traffic patterns. 

%% file: sections/appendix.tex
\appendices

\section{Measurement Setup for Agents} \label{appen:measurement} 

Figure~\ref{fig:task_prompt} shows the task execution prompt for fingerprint measurement of web agents as discussed in Section~\ref{sec:measurement-prompt}. 

\begin{figure*}[ht]
\begin{tcolorbox}[colback=gray!5!white, colframe=gray!75!black, title=Task Instruction for Autonomous Web Interaction]
\small
\begin{alltt}
You are an autonomous web interaction agent.

Your task is to visit all seven URLs exactly once, interact with each page in the way 
you judge most natural, and then move on to the next URL.

Use your own judgment to decide what the page is asking for and what short interaction 
sequence is appropriate.

Target URLs, in order:
1. https://<testbed domain>/page-01.html
2. https://<testbed domain>/page-02.html
3. https://<testbed domain>/page-03.html
4. https://<testbed domain>/page-04.html
5. https://<testbed domain>/page-05.html
6. https://<testbed domain>/page-06.html
7. https://<testbed domain>/page-07.html

For each target URL:

1. Navigate directly to the exact URL.

2. Once the page loads, inspect the page and decide what a reasonable user would 
naturally do.

3. Perform one short natural interaction sequence.
- A sequence may contain multiple low-level browser actions if they naturally belong 
together.
- For example, typing into a field and pressing a nearby submit button may count as one 
sequence.
- Choose synthetic input values yourself.
- Do not use real personal information.
- Do not ask the user what value to enter.

4. After you perform any meaningful page interaction sequence, the current URL is 
considered complete. Immediately navigate directly to the next target URL.

5. If you are unsure what to do next, do not ask the user. Instead, navigate directly to 
the next unvisited target URL.

6. Do not retry, do not optimize, and do not attempt alternative strategies on the same 
page.

After all 7 URLs have been visited, return a concise page-by-page summary.

Important:
- Never ask the user what to do next.
- Never ask for confirmation.
- Never wait for additional instructions.
- Act naturally, but keep moving.
- Your job is not to fully solve each page.
- Your job is to perform one natural interaction per page and proceed.
\end{alltt}
\end{tcolorbox}
\caption{\textbf{Task instruction for autonomous web interaction.} It guides the agent to sequentially visit each target URL, inspect the page, perform one natural interaction with synthetic inputs, and immediately proceed to the next page.}
\label{fig:task_prompt}
\end{figure*}

\section{User Behavior Testbed Experiments}

\begin{figure}[t]
    \centering
    \includegraphics[width=0.9\columnwidth]{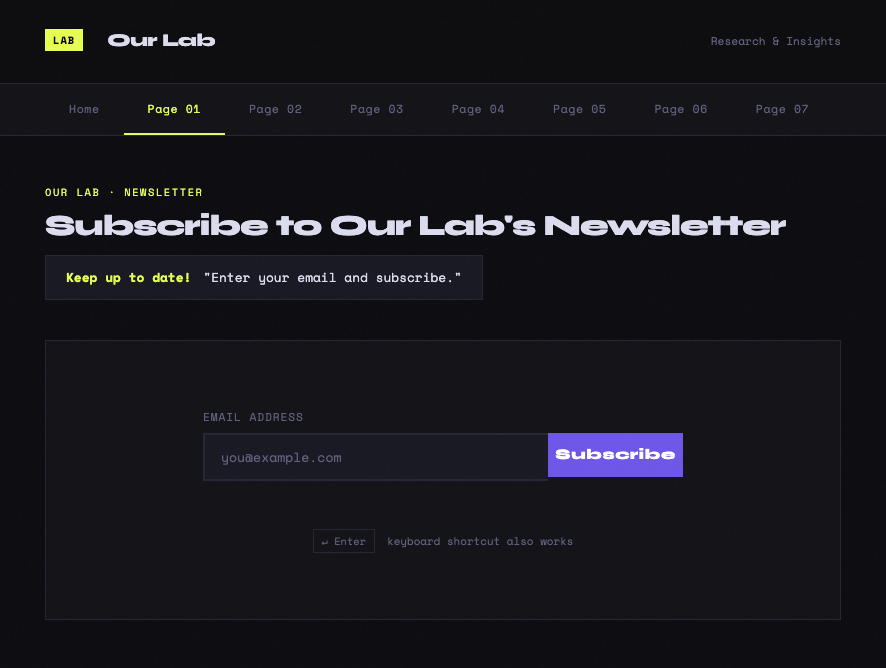}
    \caption{Scenario 1, version 1.}
    \label{fig:s1-v1}
\end{figure}

\begin{figure}[t]
    \centering
    \includegraphics[width=0.9\columnwidth]{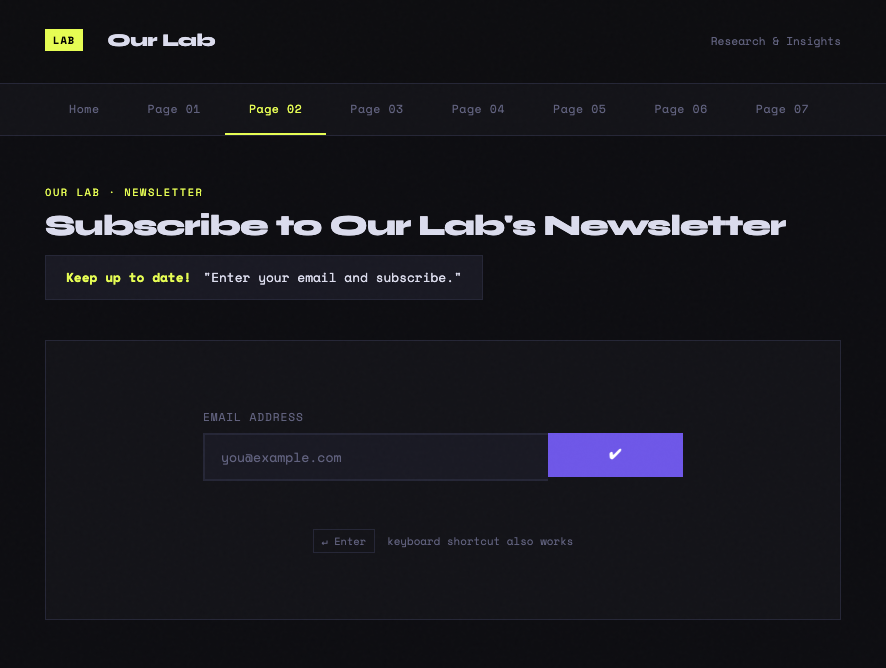}
    \caption{Scenario 1, version 2.}
    \label{fig:s1-v2}
\end{figure}

\begin{figure}[t]
    \centering
    \includegraphics[width=0.9\columnwidth]{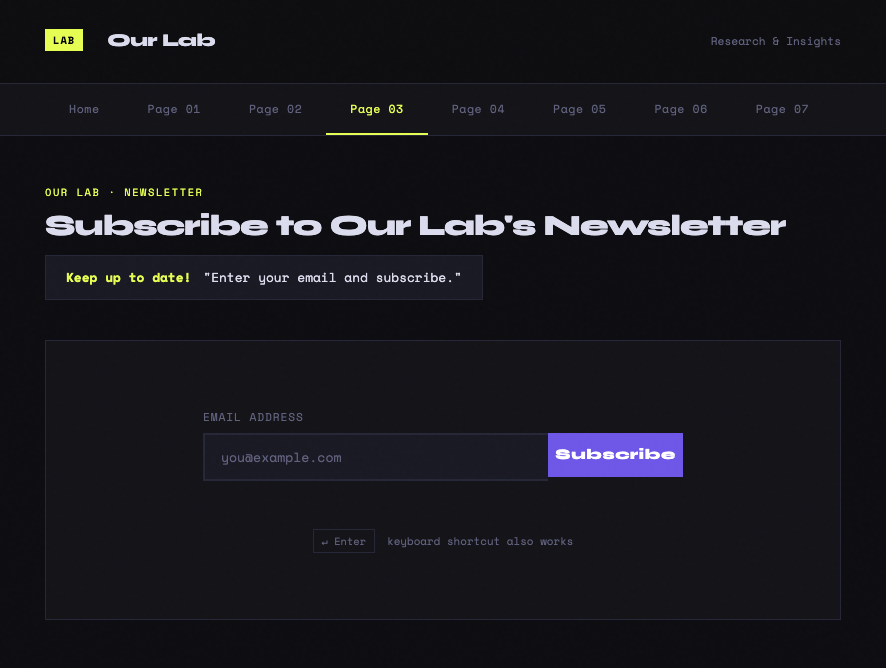}
    \caption{Scenario 1, version 3.}
    \label{fig:s1-v3}
\end{figure}

\begin{figure}[t]
    \centering
    \includegraphics[width=0.9\columnwidth]{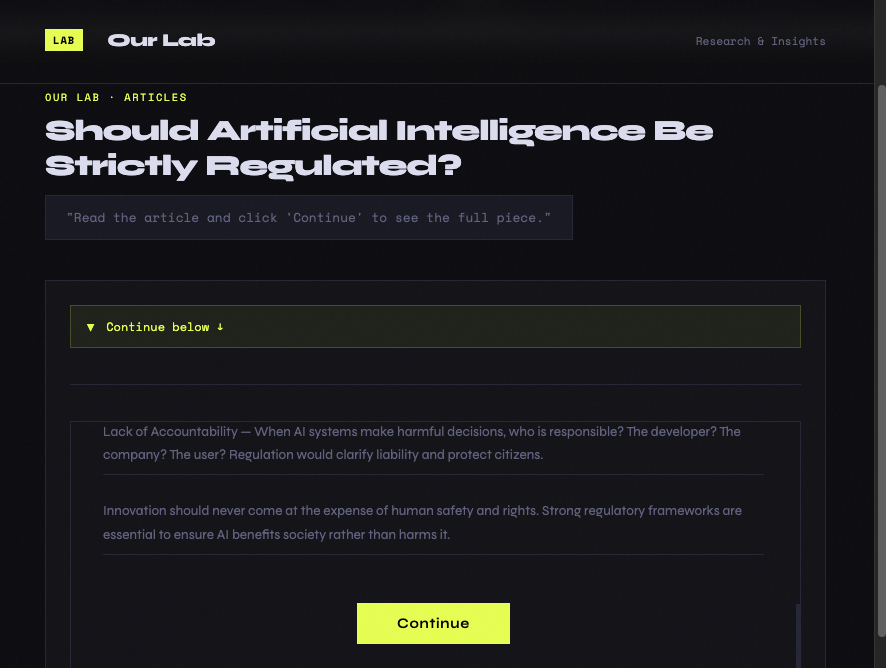}
    \caption{Scenario 2.}
    \label{fig:s2}
\end{figure}

\begin{figure}[t]
    \centering
    \includegraphics[width=0.9\columnwidth]{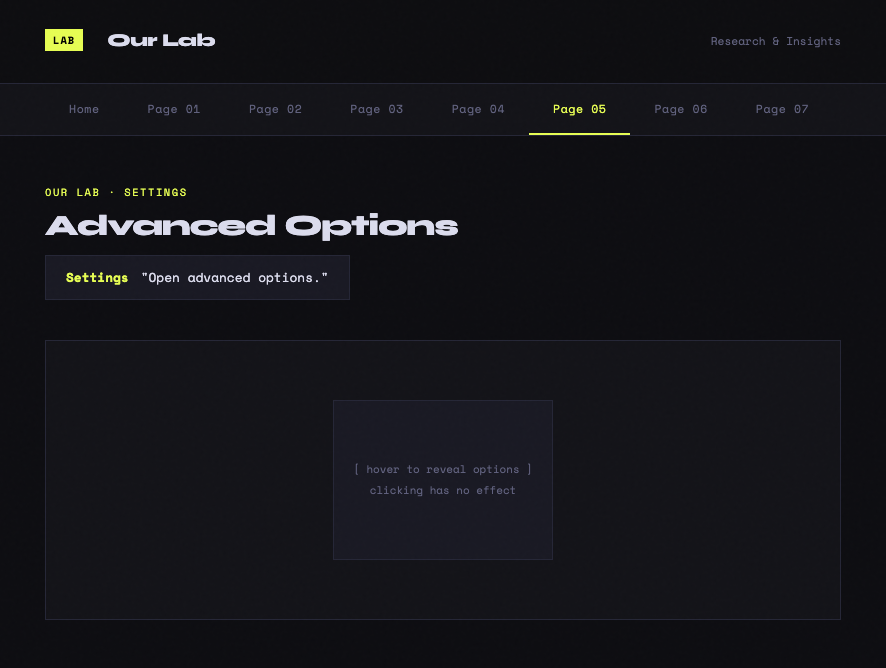}
    \caption{Scenario 3.}
    \label{fig:s3}
\end{figure}

\begin{figure}[t]
    \centering
    \includegraphics[width=0.9\columnwidth]{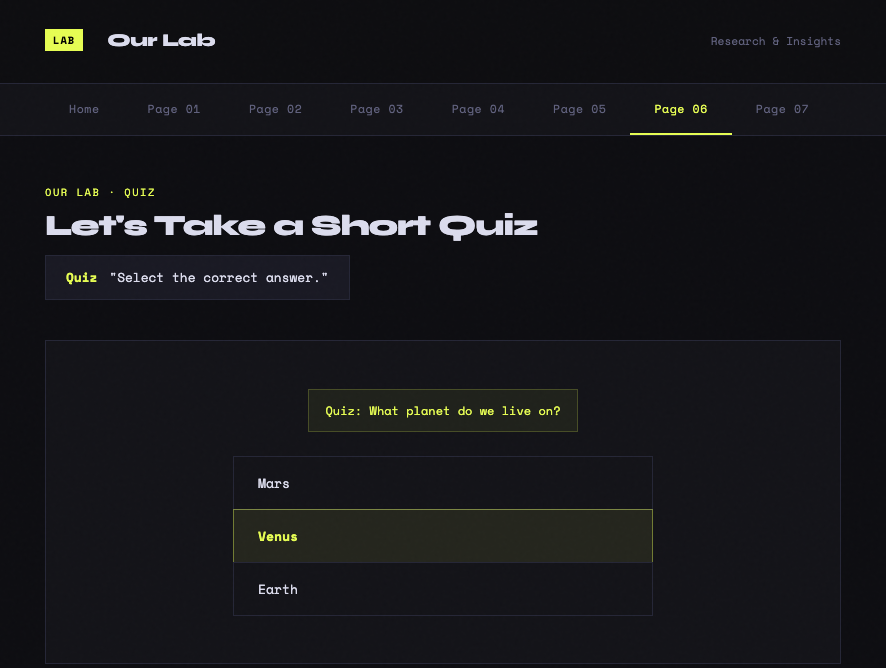}
    \caption{Scenario 4.}
    \label{fig:s4}
\end{figure}

\begin{figure}[t]
    \centering
    \includegraphics[width=0.9\columnwidth]{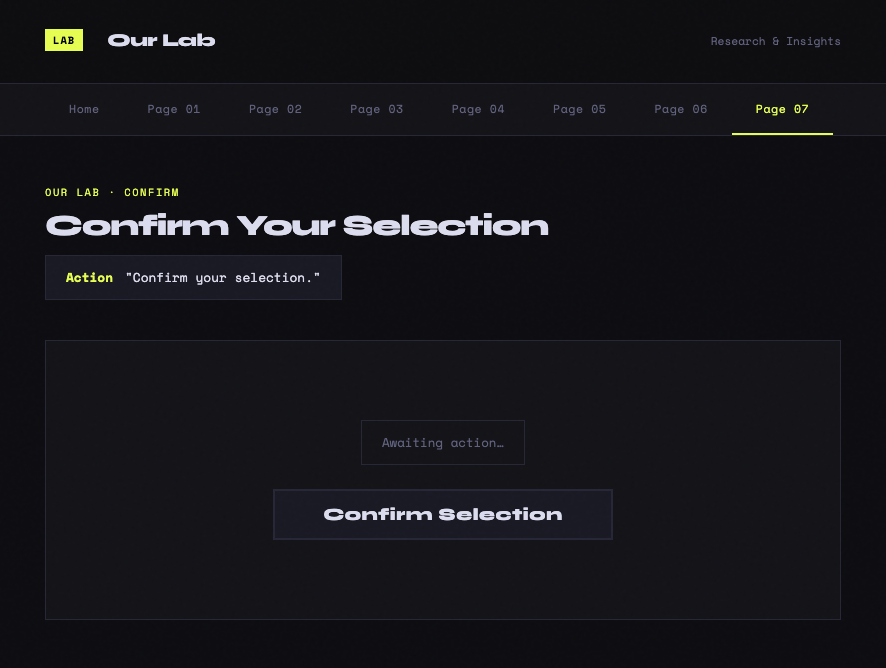}
    \caption{Scenario 5.}
    \label{fig:s5}
\end{figure}

Figures~\ref{fig:s1-v1} to~\ref {fig:s5} illustrate the actual deployed UX scenarios as we addressed in Section~\ref{sec:ux-task}.


%% file: main.bib
@inproceedings{houmansadr2013parrot,
  title={The parrot is dead: Observing unobservable network communications},
  author={Houmansadr, Amir and Brubaker, Chad and Shmatikov, Vitaly},
  booktitle={2013 IEEE Symposium on Security and Privacy},
  pages={65--79},
  year={2013},
  organization={IEEE}
}

@article{guan2026componentbench,
  title={ComponentBench: Diagnosing Component-Level Failures in Computer-Use Agents},
  author={Guan, Tianchen and Lin, Xinlei and Cheng-Yue, Royce and Wang, Xiangjun and Zhou, Shuyan},
  journal={arXiv preprint arXiv:2608.18307},
  year={2026}
}

@article{pugliese2020long,
  title={Long-term observation on browser fingerprinting: Users’ trackability and perspective},
  author={Pugliese, Gaston and Riess, Christian and Gassmann, Freya and Benenson, Zinaida},
  journal={Proceedings on Privacy Enhancing Technologies},
  year={2020}
}

@inproceedings{eckersley2010unique,
  title={How unique is your web browser?},
  author={Eckersley, Peter},
  booktitle={International Symposium on Privacy Enhancing Technologies Symposium},
  pages={1--18},
  year={2010},
  organization={Springer}
}

@inproceedings{vastel2018fp,
  title={Fp-stalker: Tracking browser fingerprint evolutions},
  author={Vastel, Antoine and Laperdrix, Pierre and Rudametkin, Walter and Rouvoy, Romain},
  booktitle={2018 IEEE Symposium on Security and Privacy (SP)},
  pages={728--741},
  year={2018},
  organization={IEEE}
}

@inproceedings{sun2002statistical,
  title={Statistical identification of encrypted web browsing traffic},
  author={Sun, Qixiang and Simon, Daniel R and Wang, Yi-Min and Russell, Wilf and Padmanabhan, Venkata N and Qiu, Lili},
  booktitle={Proceedings 2002 IEEE Symposium on Security and Privacy},
  pages={19--30},
  year={2002},
  organization={IEEE}
}

@inproceedings{hayes2016k,
  title={k-fingerprinting: A robust scalable website fingerprinting technique},
  author={Hayes, Jamie and Danezis, George},
  booktitle={25th USENIX Security Symposium (USENIX Security 16)},
  pages={1187--1203},
  year={2016}
}

@inproceedings{panchenko2016website,
  title={Website fingerprinting at internet scale.},
  author={Panchenko, Andriy and Lanze, Fabian and Pennekamp, Jan and Engel, Thomas and Zinnen, Andreas and Henze, Martin and Wehrle, Klaus},
  booktitle={NDSS},
  volume={1},
  pages={23477},
  year={2016}
}

@misc{hbsWhosAdopting,
	author = {Jeremy Yang},
	title = {Who’s Adopting AI Agents—and What They’re Actually Doing With Them},
	howpublished = {\url{https://www.library.hbs.edu/working-knowledge/whos-adopting-ai-agents-and-what-theyre-actually-doing-with-them}},
	year = {2026},
	note = {[Accessed 12-06-2026]},
}

@inproceedings{vassio2017users,
  title={Users' fingerprinting techniques from TCP traffic},
  author={Vassio, Luca and Giordano, Danilo and Trevisan, Martino and Mellia, Marco and da Silva, Ana Paula Couto},
  booktitle={Proceedings of the Workshop on Big Data Analytics and Machine Learning for Data Communication Networks},
  pages={49--54},
  year={2017}
}

@inproceedings{kirchler2016tracked,
  title={Tracked without a trace: linking sessions of users by unsupervised learning of patterns in their DNS traffic},
  author={Kirchler, Matthias and Herrmann, Dominik and Lindemann, Jens and Kloft, Marius},
  booktitle={Proceedings of the 2016 ACM workshop on artificial intelligence and security},
  pages={23--34},
  year={2016}
}

@article{gu2017novel,
  title={A novel attack to track users based on the behavior patterns},
  author={Gu, Xiaodan and Yang, Ming and Shi, Congcong and Ling, Zhen and Luo, Junzhou},
  journal={Concurrency and Computation: Practice and Experience},
  volume={29},
  number={6},
  pages={e3891},
  year={2017},
  publisher={Wiley Online Library}
}

@inproceedings{deusser2020browsing,
  title={Browsing unicity: On the limits of anonymizing web tracking data},
  author={Deu{\ss}er, Clemens and Passmann, Steffen and Strufe, Thorsten},
  booktitle={2020 IEEE Symposium on Security and Privacy (SP)},
  pages={777--790},
  year={2020},
  organization={IEEE}
}

@inproceedings{obendorf2007web,
  title={Web page revisitation revisited: implications of a long-term click-stream study of browser usage},
  author={Obendorf, Hartmut and Weinreich, Harald and Herder, Eelco and Mayer, Matthias},
  booktitle={Proceedings of the SIGCHI conference on Human factors in computing systems},
  pages={597--606},
  year={2007}
}

@article{yang2010web,
  title={Web user behavioral profiling for user identification},
  author={Yang, Yinghui Catherine},
  journal={Decision Support Systems},
  volume={49},
  number={3},
  pages={261--271},
  year={2010},
  publisher={Elsevier}
}

@article{herrmann2013behavior,
  title={Behavior-based tracking: Exploiting characteristic patterns in DNS traffic},
  author={Herrmann, Dominik and Banse, Christian and Federrath, Hannes},
  journal={Computers \& Security},
  volume={39},
  pages={17--33},
  year={2013},
  publisher={Elsevier}
}

@inproceedings{banse2012tracking,
  title={Tracking users on the internet with behavioral patterns: Evaluation of its practical feasibility},
  author={Banse, Christian and Herrmann, Dominik and Federrath, Hannes},
  booktitle={IFIP International Information Security Conference},
  pages={235--248},
  year={2012},
  organization={Springer}
}

@article{crichton2025rethinking,
  title={Rethinking Fingerprinting: An Assessment of Behavior-based Methods at Scale and Implications for Web Tracking},
  author={Crichton, Kyle and Cranor, Lorrie Faith and Christin, Nicolas},
  journal={Proceedings on Privacy Enhancing Technologies},
  year={2025}
}

@misc{letsencryptAboutLets,
	author = {Let’s Encrypt },
	title = {{A}bout {L}et's {E}ncrypt},
	howpublished = {\url{https://letsencrypt.org/about/}},
	year = {2021},
	note = {[Accessed 11-06-2026]},
}

@misc{scrapyScrapyOpen,
	author = {Scrapy},
	title = {{S}crapy — open source web scraping framework for {P}ython},
	howpublished = {\url{https://www.scrapy.org/}},
	year = {},
	note = {[Accessed 11-06-2026]},
}

@misc{githubGitHubInternetarchiveheritrix3,
	author = {internetarchive/heritrix3},
	title = {Heritrix},
	howpublished = {\url{https://github.com/internetarchive/heritrix3}},
	year = {},
	note = {[Accessed 11-06-2026]},
}

@misc{techstackupsBrowserClaude,
	author = {James Daniel Whitford},
	title = {{B}rowser {U}se vs {C}laude {C}omputer {U}se: {D}{O}{M} vs {V}ision},
	howpublished = {\url{https://techstackups.com/comparisons/browser-use-vs-claude-computer-use/}},
	year = {2026},
	note = {[Accessed 23-10-2026]},
}

@misc{githubLogicalArchitecture,
	author = {},
	title = {Browser Use - {L}ogical {A}rchitecture},
	howpublished = {\url{https://github.com/swisherae/browser-use/wiki/{L}ogical-{A}rchitecture}},
	year = {},
	note = {[Accessed 23-10-2026]},
}

@misc{tinyfishWhatBrowser,
	author = {Rina Takahashi},
	title = {{W}hat {B}rowser {A}gents {S}ee {B}efore {T}hey {T}hink},
	howpublished = {\url{https://current.tinyfish.ai/issue/31/foundations/article/18062/what-browser-agents-see-before-they-think}},
	year = {2026},
	note = {[Accessed 23-10-2026]},
}

@misc{notteDoesBrowser,
	author = {Lucas Giordano},
	title = {{H}ow does a browser agent perceive a page (vision vs {D}{O}{M})?},
	howpublished = {\url{https://www.notte.cc/glossary/ai-browser-agents/how-does-a-browser-agent-perceive-a-page-vision-vs-dom}},
	year = {2026},
	note = {[Accessed 23-10-2026]},
}

@misc{how-autogen-websurfer-works,
	author = {},
	title = {AutoGen WebSurfer API reference - autogen\_ext.agents.web\_surfer},
	howpublished = {\url{https://microsoft.github.io/autogen/stable/reference/python/autogen\_ext.agents.web\_surfer.html}},
	year = {},
	note = {[Accessed 23-10-2026]},
}

@misc{how-skyvern-works,
	author = {Skyvern},
	title = {How Skyvern works},
	howpublished = {\url{https://www.skyvern.com/docs/concepts.md\#how-skyvern-works}},
	year = {},
	note = {[Accessed 23-10-2026]},
}

@misc{chatgpt-work,
	author = {OepnAI},
	title = {ChatGPT Work and Codex},
	howpublished = {\url{https://help.openai.com/en/articles/20001275-chatgpt-work-and-codex}},
	year = {2026},
	note = {[Accessed 22-10-2026]},
}

@misc{apacheApacheNutch,
	author = {Apache Nutch Project Management Committee},
	title = {{A}pache {N}utch},
	howpublished = {\url{https://nutch.apache.org/}},
	year = {},
	note = {[Accessed 11-06-2026]},
}

@misc{wang2026fpagentfingerprintingaibrowsing,
      title={FP-Agent: Fingerprinting AI Browsing Agents}, 
      author={Ethan Wang and Zubair Shafiq and Yash Vekaria},
      year={2026},
      eprint={2605.01247},
      archivePrefix={arXiv},
      primaryClass={cs.CR},
      url={https://arxiv.org/abs/2605.01247}, 
}

@misc{webfingerprint2026,
	author = {{Fingerprint}},
	title = {{F}ingerprint {L}aunches {A}utomation {I}ntelligence {A}{P}{I} and {A}{I} {A}ssistant {D}etection, {D}elivering the {I}ndustry's {M}ost {C}omplete {V}iew of {A}{I} {T}raffic},
	howpublished = {\url{https://www.businesswire.com/news/home/20260601158287/en/Fingerprint-Launches-Automation-Intelligence-API-and-AI-Assistant-Detection-Delivering-the-Industrys-Most-Complete-View-of-AI-Traffic}},
	year = {2026},
	note = {[Accessed 11-06-2026]},
}

@misc{rfc9218,
    series =    {Request for Comments},
    number =    9218,
    howpublished =  {RFC 9218},
    publisher = {RFC Editor},
    doi =       {10.17487/RFC9218},
    url =       {https://www.rfc-editor.org/info/rfc9218},
    author =    {Kazuho Oku and Lucas Pardue},
    title =     {{Extensible Prioritization Scheme for HTTP}},
    pagetotal = 21,
    year =      2022,
    month =     jun,
    note =      {[Accessed 27-05-2026]},
}

@misc{rfc7540,
    series =    {Request for Comments},
    number =    7540,
    howpublished =  {RFC 7540},
    publisher = {RFC Editor},
    doi =       {10.17487/RFC7540},
    url =       {https://www.rfc-editor.org/info/rfc7540},
    author =    {M. Belshe and R. Peon and M. Thomson},
    title =     {{Hypertext Transfer Protocol Version 2 ({HTTP}/2)}},
    year =      2015,
    month =     may,
    note =      {Obsoleted by RFC 9113. [Accessed 27-05-2026]},
}

@misc{cloudflareTrappingMisbehaving,
	author = {Reid Tatoris and Harsh Saxena and Luis Miglietti},
	title = {{T}rapping misbehaving bots in an {A}{I} {L}abyrinth --- blog.cloudflare.com},
	howpublished = {\url{https://blog.cloudflare.com/ai-labyrinth/}},
	year = {2025},
	note = {[Accessed 02-05-2026]},
}

@inproceedings{10.1145/3730567.3732913,
author = {Liu, Enze and Luo, Elisa and Shan, Shawn and Voelker, Geoffrey M. and Zhao, Ben Y. and Savage, Stefan},
title = {Somesite I Used To Crawl: Awareness, Agency and Efficacy in Protecting Content Creators From {AI} Crawlers},
year = {2025},
isbn = {9798400718601},
publisher = {Association for Computing Machinery},
address = {New York, NY, USA},
url = {https://doi.org/10.1145/3730567.3732913},
doi = {10.1145/3730567.3732913},
booktitle = {Proceedings of the 2025 ACM Internet Measurement Conference},
pages = {78–99},
numpages = {22},
location = {Madison, WI, USA},
series = {IMC '25}
}

@inproceedings{10.1145/3730567.3764471,
author = {Kim, Taein and Bock, Karstan and Luo, Claire and Liswood, Amanda and Poroslay, Chloe and Wenger, Emily},
title = {Scrapers Selectively Respect robots.txt Directives: Evidence From a Large-Scale Empirical Study},
year = {2025},
isbn = {9798400718601},
publisher = {Association for Computing Machinery},
address = {New York, NY, USA},
url = {https://doi.org/10.1145/3730567.3764471},
doi = {10.1145/3730567.3764471},
booktitle = {Proceedings of the 2025 ACM Internet Measurement Conference},
pages = {541–557},
numpages = {17},
location = {Madison, WI, USA},
series = {IMC '25}
}

@misc{rfc9309,
    series =    {Request for Comments},
    number =    9309,
    howpublished =  {RFC 9309},
    publisher = {RFC Editor},
    doi =       {10.17487/RFC9309},
    url =       {https://www.rfc-editor.org/info/rfc9309},
    author =    {Martijn Koster and Gary Illyes and Henner Zeller and Lizzi Sassman},
    title =     {{Robots Exclusion Protocol}},
    pagetotal = 12,
    year =      2022,
    month =     sep,
}

@misc{masterofcode150Agent,
	author = {Ivan Pohrebniyak},
	title = {150+ AI Agents Statistics: What Business Leaders Are Betting On in 2026},
	howpublished = {\url{https://masterofcode.com/blog/ai-agent-statistics}},
	year = {2026},
	note = {[Accessed 02-05-2026]},
}

@misc{thebusinessresearchcompanyBusinessResearch,
	author = {The Business Research Company},
	title = {{A}{I} {A}gents {M}arket {S}ize {R}eport 2026, {G}rowth, {A}nalysis {A}nd {F}orecast --- thebusinessresearchcompany.com},
	howpublished = {\url{https://www.thebusinessresearchcompany.com/report/ai-agents-global-market-report}},
	year = {2026},
	note = {[Accessed 02-05-2026]},
}

@article{akkil2026emergence,
  title={Emergence {WebVoyager}: Toward Consistent and Transparent Evaluation of {(Web) Agents} in The Wild},
  author={Akkil, Deepak and Allaham, Mowafak and Raj, Amal and Abuelsaad, Tamer and Kokku, Ravi},
  journal={arXiv preprint arXiv:2603.29020},
  year={2026}
}

@misc{cloudflarePerplexityUsing,
	author = {Gabriel Corral and Vaibhav Singhal and Brian Mitchell and Reid Tatoris},
	title = {{P}erplexity is using stealth, undeclared crawlers to evade website no-crawl directives --- blog.cloudflare.com},
	howpublished = {\url{https://blog.cloudflare.com/perplexity-is-using-stealth-undeclared-crawlers-to-evade-website-no-crawl-directives/}},
	year = {2025},
	note = {[Accessed 21-04-2026]},
}

@inproceedings{10.1145/3719027.3765063,
author = {Cui, Jian and Zha, Mingming and Wang, XiaoFeng and Liao, Xiaojing},
title = {The Odyssey of robots.txt Governance: Measuring Convention Implications of Web Bots in {Large Language Model} Services},
year = {2025},
isbn = {9798400715259},
publisher = {Association for Computing Machinery},
address = {New York, NY, USA},
url = {https://doi.org/10.1145/3719027.3765063},
doi = {10.1145/3719027.3765063},
booktitle = {Proceedings of the 2025 ACM SIGSAC Conference on Computer and Communications Security},
pages = {21–35},
numpages = {15},
location = {Taipei, Taiwan},
series = {CCS '25}
}

@misc{browser_use2024,
  author = {Müller, Magnus and Žunič, Gregor},
  title = {{Browser Use}: Enable {AI} to control your browser},
  year         = {2024},
  publisher = {GitHub},
  howpublished = {\url{https://github.com/browser-use/browser-use}}
}

@misc{openai2025introducingdeepresearch,
  author       = {{OpenAI}},
  title        = {{Introducing Deep Research}},
  howpublished = {\url{https://openai.com/index/introducing-deep-research/}},
  year         = {2025},
}

@misc{google2025geminideepresearch,
  author       = {{Google}},
  title        = {{Gemini Deep Research --- your personal research assistant}},
  howpublished = {\url{https://gemini.google/overview/deep-research/}},
  year         = {2025},
  note         = {Accessed: 2025-07-16},
}

@misc{anthropic2026computeruse,
  title        = {Computer use tool},
  author       = {{Anthropic}},
  year         = {2026},
  howpublished = {\url{https://platform.claude.com/docs/en/agents-and-tools/tool-use/computer-use-tool}},
  note         = {Claude API documentation. Accessed: 2026-05-01}
}

@misc{skyvern2026,
  author = {{Skyvern-AI}},
  title = {Skyvern: Automate browser-based workflows with {AI}},
  year = {2026},
  publisher = {GitHub},
  howpublished = {\url{https://github.com/Skyvern-AI/skyvern}},
  note = {Accessed: 2026-05-01}
}

@misc{cursor2026,
  title        = {Build Software with {AI} Agents},
  author       = {{Cursor}},
  year         = {2026},
  howpublished = {\url{https://cursor.com/product}},
  note         = {Accessed: 2026-05-01}
}

@inproceedings{yang2024sweagent,
  title={{SWE}-agent: Agent-Computer Interfaces Enable Automated Software Engineering},
  author={John Yang and Carlos E Jimenez and Alexander Wettig and Kilian Lieret and Shunyu Yao and Karthik R Narasimhan and Ofir Press},
  booktitle={NeurIPS},
  year={2024},
  pages={50528--50652},
  url={https://arxiv.org/abs/2405.15793}
}

@misc{openai2026codex,
  title        = {Codex},
  author       = {{OpenAI}},
  year         = {2026},
  howpublished = {\url{https://developers.openai.com/codex}},
  note         = {OpenAI Developers documentation. Accessed: 2026-05-01}
}

@misc{skyvern2025readsweb,
  title        = {How {Skyvern} Reads and Understands the Web},
  author       = {Suchintan Singh},
  year         = {2025},
  month        = jul,
  day          = {16},
  howpublished = {\url{https://www.skyvern.com/blog/how-skyvern-reads-and-understands-the-web/}},
  note         = {Accessed: 2026-05-01}
}

@misc{openai2025cua,
  title        = {Computer-Using Agent},
  author       = {{OpenAI}},
  year         = {2025},
  month        = jan,
  day          = {23},
  howpublished = {\url{https://openai.com/index/computer-using-agent/}},
  note         = {Accessed: 2026-05-01}
}

@misc{microsoft2025omnitool,
  title        = {{OmniTool}: Computer Use with {OmniParser}},
  author       = {{Microsoft}},
  year         = {2025},
  howpublished = {\url{https://github.com/microsoft/OmniParser/blob/master/omnitool/readme.md}},
  note         = {Accessed: 2026-05-01}
}

@misc{anthropic2026claudecode,
  title        = {{Claude} {Code} Overview},
  author       = {{Anthropic}},
  year         = {2026},
  howpublished = {\url{https://code.claude.com/docs/en/overview}},
  note         = {Claude Code documentation. Accessed: 2026-05-01}
}

@misc{cognition2024devin,
  title        = {Introducing Devin, the First {AI} Software Engineer},
  author       = {{Cognition}},
  year         = {2024},
  month        = mar,
  day          = {12},
  howpublished = {\url{https://cognition.ai/blog/introducing-devin}},
  note         = {Accessed: 2026-05-01}
}

@misc{openai2025operator,
  title        = {Introducing {Operator}},
  author       = {{OpenAI}},
  year         = {2025},
  month        = jan,
  day          = {23},
  howpublished = {\url{https://openai.com/index/introducing-operator/}},
  note         = {Accessed: 2026-05-01}
}

@misc{google2026computeruse,
  title        = {Computer Use},
  author       = {{Google}},
  year         = {2026},
  howpublished = {\url{https://ai.google.dev/gemini-api/docs/computer-use}},
  note         = {Gemini API documentation. Last updated: 2026-04-28. Accessed: 2026-05-01}
}

@misc{elovic2023gptresearcher,
  author       = {Assaf Elovic},
  title        = {{GPT} Researcher},
  date         = {2023-07-25},
  url          = {https://gptr.dev},
  note         = {Code repository: \url{https://github.com/assafelovic/gpt-researcher}}
}

@article{stockl2025ai,
  title={Are {AI} Agents interacting with Online Ads?},
  author={St{\"o}ckl, Andreas and Nitu, Joel},
  journal={arXiv preprint arXiv:2504.07112},
  year={2025}
}

@inproceedings{yao2023react,
  title={{ReAct}: Synergizing Reasoning and Acting in Language Models},
  author={Yao, Shunyu and Zhao, Jeffrey and Yu, Dian and Du, Nan and Shafran, Izhak and Narasimhan, Karthik and Cao, Yuan},
  booktitle={ICLR},
  year={2023}
}

@article{wang2023voyager,
  title={Voyager: An Open-Ended Embodied Agent with {Large Language Models}},
  author={Wang, Guanzhi and Xie, Yuqi and Jiang, Yunfan and Mandlekar, Ajay and Xiao, Chaowei and Zhu, Yuke and Fan, Linxi and Anandkumar, Anima},
  journal={Transactions on Machine Learning Research},
  year={2024},
  url={https://openreview.net/forum?id=ehfRiF0R3a},
  eprint={2305.16291},
  archivePrefix={arXiv}
}

@inproceedings{bazinska2026breaking,
  title={Breaking Agent Backbones: Evaluating the Security of Backbone {LLMs} in {AI} Agents},
  author={Bazinska, Julia and Mathys, Max and Casucci, Francesco and Rojas-Carulla, Mateo and Davies, Xander and Souly, Alexandra and Pfister, Niklas},
  booktitle={The Fourteenth International Conference on Learning Representations},
  year={2026}
}

@misc{althouse2019ja3,
  author       = {John Althouse and Jeff Atkinson and Josh Atkins},
  title        = {{TLS} Fingerprinting with {JA3} and {JA3S}},
  howpublished = {Salesforce Engineering Blog},
  year         = {2019},
  month        = jan,
  note         = {\url{https://engineering.salesforce.com/tls-fingerprinting-with-ja3-and-ja3s-247362855967/}}
}

@misc{althouse2023ja4,
  author       = {John Althouse},
  title        = {{JA4+} Network Fingerprinting},
  howpublished = {FoxIO Blog},
  year         = {2023},
  month        = sep,
  note         = {\url{https://foxio.io/blog/ja4-network-fingerprinting}}
}

@misc{segal2017http2,
  author       = {Ory Segal and Aharon Fridman and Elad Shuster},
  title        = {Passive Fingerprinting of {HTTP/2} Clients},
  howpublished = {Akamai Technologies White Paper},
  year         = {2017},
  note         = {Presented at Black Hat Europe 2017. \url{https://www.blackhat.com/docs/eu-17/materials/eu-17-Shuster-Passive-Fingerprinting-Of-HTTP2-Clients-wp.pdf}}
}

@article{husak2016https,
  author  = {Hus{\'a}k, Martin and {\v{C}}erm{\'a}k, Milan and Jirs{\'i}k, Tom{\'a}{\v{s}} and {\v{C}}eleda, Pavel},
  title   = {{HTTPS} traffic analysis and client identification using passive {SSL/TLS} fingerprinting},
  journal = {EURASIP Journal on Information Security},
  year    = {2016},
  volume  = {2016},
  pages   = {6},
  doi     = {10.1186/s13635-016-0030-7}
}

@inproceedings{frolov2019tls,
  author    = {Frolov, Sergey and Wustrow, Eric},
  title     = {The use of {TLS} in Censorship Circumvention},
  booktitle = {Proceedings of the 26th Annual Network and Distributed System Security Symposium (NDSS)},
  year      = {2019},
  publisher = {Internet Society},
  doi       = {10.14722/ndss.2019.23511}
}

@inproceedings{anderson2019tls,
  author    = {Anderson, Blake and McGrew, David A.},
  title     = {{TLS} Beyond the Browser: Combining End Host and Network Data to Understand Application Behavior},
  booktitle = {Proceedings of the Internet Measurement Conference (IMC '19)},
  year      = {2019},
  pages     = {379--392},
  publisher = {ACM},
  doi       = {10.1145/3355369.3355601}
}

@inproceedings{li2021goodbot,
  author    = {Li, Xigao and Amin Azad, Babak and Rahmati, Amir and Nikiforakis, Nick},
  title     = {Good Bot, Bad Bot: Characterizing Automated Browsing Activity},
  booktitle = {2021 IEEE Symposium on Security and Privacy (SP)},
  year      = {2021},
  pages     = {1589--1605},
  publisher = {IEEE},
  doi       = {10.1109/SP40001.2021.00079}
}

@misc{jarad2026handshakes,
  title         = {When Handshakes Tell the Truth: Detecting Web Bad Bots via {TLS} Fingerprints},
  author        = {Jarad, Ghalia and Bicakci, Kemal},
  year          = {2026},
  eprint        = {2602.09606},
  archivePrefix = {arXiv},
  primaryClass  = {cs.CR},
  url           = {https://arxiv.org/abs/2602.09606}
}

@misc{zhang2025agentprint,
  title         = {Exposing {LLM} User Privacy via Traffic Fingerprint Analysis: A Study of Privacy Risks in {LLM} Agent Interactions},
  author        = {Zhang, Yixiang and Deng, Xinhao and Gu, Zhongyi and Chen, Yihao and Xu, Ke and Li, Qi and Wu, Jianping},
  year          = {2025},
  eprint        = {2510.07176},
  archivePrefix = {arXiv},
  primaryClass  = {cs.CR},
  url           = {https://arxiv.org/abs/2510.07176}
}

@inproceedings{wu2023autogen,
  title={{AutoGen}: Enabling Next-Gen {LLM} Applications via Multi-Agent Conversations},
  author={Wu, Qingyun and Bansal, Gagan and Zhang, Jieyu and Wu, Yiran and Li, Beibin and Zhu, Erkang and Jiang, Li and Zhang, Xiaoyun and Zhang, Shaokun and Liu, Jiale and Awadallah, Ahmed Hassan and White, Ryen W and Burger, Doug and Wang, Chi},
  booktitle={COLM},
  year=2024
}
